\pdfoutput=1

\documentclass[a4paper,fleqn]{cas-sc}

\usepackage[numbers]{natbib}

\usepackage{hyperref}
\usepackage{amsthm}
\usepackage[ruled, vlined, linesnumbered]{algorithm2e}
\usepackage{adjustbox}

\usepackage{enumitem}

\usepackage{tikz}
\usetikzlibrary{arrows,automata,backgrounds,calc,chains,external,fit,graphs,patterns,positioning,shapes,shapes.geometric}

\newcommand\InvAAST[0]{\textsc{Inv\-AAST\-Cluster}\xspace}
\newcommand\IPAs[0]{\textsc{IPAs}\xspace}
\newcommand\IPA[0]{\textsc{IPA}\xspace}
\newcommand\AASTs[0]{\textsc{AASTs}\xspace}
\newcommand\AAST[0]{\textsc{AAST}\xspace}
\newcommand\AST[0]{\textsc{AST}\xspace}
\newcommand\ASTs[0]{\textsc{ASTs}\xspace}
\newcommand{\benchmark}[0]{\textsc{C-Pack-IPAs}\xspace}

\newcommand\Daikon[0]{\textsc{Daikon}\xspace}
\newcommand\Clara[0]{\textsc{Clara}\xspace}

\theoremstyle{definition}
\newtheorem{definition}{Definition}[section]

\newtheorem{example}{Example}

\usepackage{listings}
\usepackage{float}
\floatstyle{plaintop}
\usepackage{caption}
\usepackage[frozencache,cache dir=.]{minted}
\newenvironment{code}{\floatstyle{plaintop}%
\captionsetup{type=listing, labelfont=bf,justification=raggedright, singlelinecheck=false}}{}

\usepackage{booktabs}   
\usepackage{subcaption} 

\def\tsc#1{\csdef{#1}{\textsc{\lowercase{#1}}\xspace}}
\tsc{WGM}
\tsc{QE}

\begin{document}
\let\WriteBookmarks\relax
\def\floatpagepagefraction{1}
\def\textpagefraction{.001}

\shorttitle{InvAASTCluster: AASTs and Invariant-Based Program Clustering}    

\shortauthors{P. Orvalho, M. Janota, and V. Manquinho}

\title [mode = title]{InvAASTCluster: On Applying Invariant-Based Program Clustering to Introductory Programming Assignments}  



%

\author[1]{Pedro Orvalho}[type=author,
      style=english,
      orcid=0000-0002-7407-5967]
\cormark[1]
\ead{pmorvalho@tecnico.ulisboa.pt}

\ead[url]{https://arsr.inesc-id.pt/~pmorvalho}

\author[2]{Mikoláš Janota}[type=author,
      style=english,
      orcid=0000-0003-3487-784X]
\ead{mikolas.janota@cvut.cz}
\ead[url]{http://people.ciirc.cvut.cz/~janotmik}

\author[1]{Vasco Manquinho}[type=author,
      style=english,
      orcid=0000-0002-4205-2189]
\ead{vasco.manquinho@tecnico.ulisboa.pt}

\ead[url]{https://arsr.inesc-id.pt/~vmm}





\affiliation[1]{organization={INESC-ID, Instituto Superior Técnico, Universidade de Lisboa},
            addressline={Rua Alves Redol 9}, 
            city={Lisboa},
            postcode={1000-029}, 
            country={Portugal}}

\affiliation[2]{organization={CIIRC, Czech Technical University in Prague},
            addressline={Dejvice}, 
            city={Prague},
            postcode={16000}, 
            country={Czechia}}
            
\cortext[1]{Corresponding author}



\begin{abstract}
Due to the vast number of students enrolled in programming courses, there has been an increasing number of automated program repair techniques focused on introductory programming assignments (\IPAs).
Typically, such techniques use program clustering to take advantage of previous correct student implementations to repair a new incorrect submission. These repair techniques use clustering methods since analyzing all available correct submissions to repair a program is not feasible.
However, conventional clustering methods rely on program representations based on features such as abstract syntax trees (\ASTs), syntax, control flow, and data flow.

This paper proposes \InvAAST, a novel approach for program clustering that uses dynamically generated program invariants to cluster semantically equivalent \IPAs.
\InvAAST’s program representation uses a combination of the program's semantics, through its invariants, and its structure through its anonymized abstract syntax tree (\AASTs). Invariants denote conditions that must remain true during program execution, while \AASTs are \ASTs devoid of variable and function names, retaining only their types.
Our experiments show that the proposed program representation outperforms syntax-based representations when clustering a set of correct \IPAs. Furthermore, we integrate \InvAAST into a state-of-the-art clustering-based program repair tool. Our results show that \InvAAST advances the current state-of-the-art when used by clustering-based repair tools by repairing around 13\% more students' programs, in a shorter amount of time.
\end{abstract}

\begin{keywords}
 Program Clustering\\
 Program Invariants\\ Program Equivalence\\ Program Repair\\ 
 Programming Education\\ MOOCS
\end{keywords}

\maketitle

\section{Introduction}
\label{sec:intro}

Nowadays, thousands of students enroll every year in pro\-gra\-mming-o\-ri\-ented Massive Open Online Courses (MOOCs)~\cite{clara}. On top of that, due to the recent pandemic situation, even small-sized programming courses are being taught online. Providing feedback to novice students in introductory programming assignments (\IPAs) in these courses requires substantial effort and time by the faculty. Hence, there is an increasing need for 
automated semantic program repair frameworks~\cite{clara, semFix, directFix, searchRepair, angelix, asr-for-ITSP, gulwani-fse14, zimmerman-ase15}
capable of providing automated, comprehensive, and personalized feedback for students' incorrect solutions to programming assignments. 

Over the last few years,
several program repair tools~\cite{clara,sarfgen,semCluster-pldi19, refactory} 
have appeared that use a large number of diverse correct implementations submitted for each \IPA by previously enrolled students. 
Given an incorrect student submission, these frameworks use clustering methods to find
the most similar correct submission from previous years to provide a minimal set of repairs to the student.
Using the same reference implementation to fix all incorrect programs can potentially generate a large set of repairs. On the other hand, having a similar correct implementation allows computing a smaller set of repairs.
However, comparing on the fly all previous correct student submissions against a given incorrect submission is not feasible.

To tackle this problem, different program clustering approaches have been used in program repair tools, which enable focusing only on the representatives of each cluster.
\Clara~\cite{clara} clusters the correct programs based on their dynamic equivalence~\cite{dynamic-equivalence} and control flow, i.e., the order in which program statements, instructions, and function calls are executed. \textsc{Sarfgen}~\cite{sarfgen} computes program representations based on each program's abstract syntax tree. 
\textsc{SemCluster}~\cite{semCluster-pldi19} uses each program's control and data flow. A program's data flow tracks the number of occurrences of consecutive values a variable takes during its lifetime.

The problem of program equivalence, i.e., deciding if two programs are equivalent, is 
undecidable~\cite{rice1953classes, pldi19-semantic-equivalence-checking,master-thesis-pedro}.
On that account, finding an adequate representation for programs that performs well on program clustering is a challenging problem.
The previously-mentioned program representations used in the field of program repair may be brittle. 
For instance, consider two programs that calculate the sum of natural numbers from 1 to a given number, one using a while-loop and the other a for-loop. Despite producing the same result, their syntactic and structural differences pose challenges for conventional program representations to recognize their semantic equivalence, as we will show in Section~\ref{sec:motivation}. To address this problem, we propose to use dynamically-generated program invariants to cluster semantically equivalent programs, overcoming some of the identified weaknesses.
A program invariant is a condition that must always be true at a given step of the program during its execution (see Section~\ref{sec:invariants}). Program invariants are usually used to assert assurances throughout a program (assertions). 

This paper proposes to leverage the information of a program's structure using its \emph{abstract syntax tree} (\AST) together with semantic information provided by its invariants.
Previous research has been conducted regarding using invariants to promote patch diversity (i.e., diversity in the set of possible repairs to a given incorrect program) on search-based program repair~\cite{ase19-weimer-understanding-patches-through-invariants, gi-icse19-invariants-diversity-search-based-repair, ibf20-yang-invariants-difference-patches}. These works use \Daikon~\cite{daikon07} to generate invariant sets for each possible patch. \Daikon is a system that infers likely dynamically generated invariants observed over several program executions. Therefore, these invariants depend on the program executions. Nevertheless, previous work~\cite{ase19-weimer-understanding-patches-through-invariants} showed promising results in using invariants to semantically cluster patches to provide the user with a semantic reason for a set of similar patches. 

This paper presents a novel approach for clustering introductory programming assignments (\IPAs) leveraging their sets of invariants. Our approach for clustering \IPAs also takes into account each program's code and \emph{anonymized abstract syntax tree (AAST)}. \AASTs are essentially \ASTs stripped of variable and function identifiers, preserving only their respective types (see Section~\ref{sec:aast}). The main contribution of this work is a vector representation of programs based on their invariants and \AASTs, bringing together their semantic and syntactic features. 
The proposed clustering technique has been implemented in a framework \InvAAST. This tool has been designed as an independent clustering tool. 
Therefore, it can be used to help evaluate students' submissions for \IPAs by clustering semantically equivalent solutions for programming exercises.
However, \InvAAST can also be easily integrated into any clustering-based program repair tool for \IPAs. Furthermore, \InvAAST can even be used in a plagiarism detection tool, like \textsc{Moss}~\cite{moss}.

\begin{figure}
    \centering
    \scalebox{0.6}{
    \includegraphics{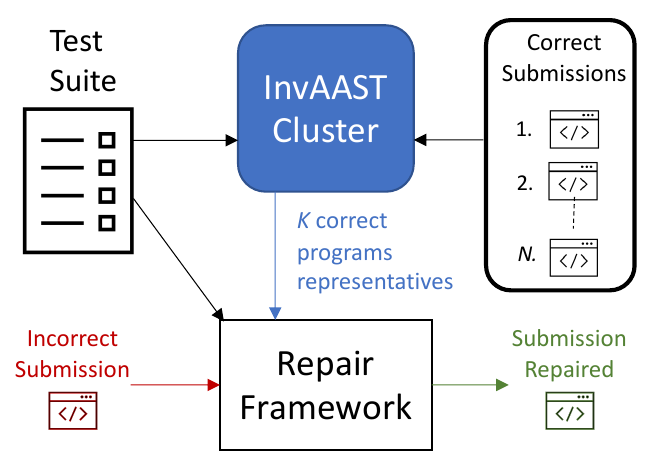}}
    \caption{Clustering-based Program Repair.}
    \label{fig:repair-process}
\end{figure}

Figure~\ref{fig:repair-process} shows the generic architecture of clustering-based program repair frameworks~\cite{clara, semCluster-pldi19,sarfgen}. These 
frameworks receive an incorrect student submission, a test suite, and a collection of $N$ correct student submissions for the same \IPA. 
For scalability concerns, these frameworks eliminate, through clustering techniques, semantically equivalent solutions, i.e., dynamically equivalent correct programs, given the provided input-output test suite. Those clustering approaches try aggregating the set of $N$~correct solutions into $K$~semantically different clusters ($N \gg K$). Finally, the repair tool uses these $K$ clusters' representatives to repair the provided incorrect student submission. 
As Figure~\ref{fig:repair-process} shows, 
\InvAAST can be used as the clustering technique of those clustering-based program repair tools.
However, some program repair tools~\cite{verifix, autograder} use a single reference implementation provided by the lecturer to repair a student's program. Typically, these tools can only use one correct implementation to repair each program.
Therefore, \InvAAST was designed to be also capable of finding on a set of correct student submissions which submission is the closest correct solution to the incorrect program. Thus, \InvAAST can suggest a specific reference implementation for each incorrect submission that may require fewer changes to fix the program.

We evaluate \InvAAST on \benchmark~\cite{C-Pack-IPAs_apr24}, a real-world student programs developed during a university introductory programming course. Experimental results show
that the proposed invariant-based representation improves upon syntax-based representations when performing program clustering. 
Additionally, we integrate \InvAAST into \Clara, a clustering-based program repair tool, in order to compare our clustering technique against \Clara's clustering method, which is the current publicly available state-of-the-art method for clustering \IPAs. 

To summarize, this paper makes the following contributions:
\begin{itemize}
	\item We propose a novel and efficient approach for clustering submissions for introductory programming assignments (\IPAs) based on the submissions' sets of invariants and \AASTs representations.
	\item We present a study showing the results of using our program clustering tool, \InvAAST, on a set of 1620 real-world \IPAs correct submissions to show the effectiveness of invariant-based program clustering.
	\item We compare \InvAAST with the clustering method used by the currently available state-of-the-art program repair tools. Experimental results show that \InvAAST outperforms state-of-the-art clustering methods, allowing clustering-based program repair tools to fix around 13\% more \IPAs in a shorter amount of time.
	\item The \InvAAST framework 
	is publicly available on GitHub at \href{https://github.com/pmorvalho/InvAASTCluster}{https://github.com/\-pm\-or\-va\-lho/\-Inv\-AAST\-Clus\-ter}.
\end{itemize}

The structure of the remainder of this paper is as follows.
First, Section~\ref{sec:motivation} illustrates the strengths of using invariants for
program representation. Section~\ref{sec:trees-and-invariants} presents important concepts used throughout this manuscript and describes how to gather and represent sets of program invariants. Section~\ref{sec:representations} discusses several program representations, including a new invariant-based program representation.
Section~\ref{sec:implementation} discusses the implementation of \InvAAST. 
Section~\ref{sec:results} presents the experimental evaluation that supports our claim that invariant-based program representations are beneficial to cluster programming assignments semantically. Finally, Section~\ref{sec:related-work} presents the related work, and the paper concludes in Section~\ref{sec:conclusion}.

\section{Motivation}
\label{sec:motivation}

Current program representations for repairing students' programming assignments leverage certain program features, such as code syntax~\cite{deepfix}, abstract syntax tree~\cite{sarfgen}, control flow~\cite{clara}, and data flow~\cite{semCluster-pldi19}, to encode each program into a vector representation. However, all of these features 
have some weaknesses when we want to cluster programs based on their semantics. 

\begin{example} 
\label{example-cycles}
Consider the following two programs written in C, that compute the sum of all the natural numbers from $1$ to a given number \texttt{n}, i.e., $\sum_{i=1}^{n}{i}$. 

\vspace{2pt}
\definecolor{mygreen}{rgb}{0,0.6,0}

\hspace*{2em}          
\begin{minipage}[t!]{0.4\linewidth}
\begin{code}
\label{prog::example_while}
\begin{minted}[escapeinside=çç,tabsize=1,obeytabs,xleftmargin=5pt,linenos]{C}
int n, sum = 0, i;
scanf("%d", &n);
i = 0;
while(i < n) {
  i++;
  sum = sum + i;
}
printf("%d\n",sum);
\end{minted}
\end{code}
\end{minipage}
\hfill
\begin{minipage}[t!]{0.45\linewidth}
\begin{code}
\label{prog::example_for}
\begin{minted}[escapeinside=çç,tabsize=1,obeytabs,xleftmargin=5pt,linenos]{C}
int j, n, s = 0;
scanf("%d", &n);

for(j = n; j > 0; j--)
{
  s = j + s;
}
printf("%d\n", s);
\end{minted}
\end{code}
\end{minipage}
\end{example}

Observe that the program on the left, in Example~\ref{example-cycles}, uses a while-loop that iterates over the natural numbers from $0$ to \texttt{n}. 
The program on the right uses a for-loop that iterates from \texttt{n} to $0$ in decreasing order. However, both programs are semantically equivalent since both have the same result. Nevertheless, if we build a program representation using the programs' syntax or abstract syntax trees, both programs will have very different representations. In terms of syntax, the names of the used variables (e.g.\ \texttt{i, j, s, sum}) and structures (e.g.\ \texttt{while}, \texttt{for}) are different. Additionally, in terms of data flow and dynamic equivalence, both programs are also different since, for example, the values assigned to the variable \texttt{i} go from $0$ to \texttt{n} in the first program while in the other the variable \texttt{j} is assigned the same values but in decreasing order. 

Consider that the variable \texttt{n} is always assigned to a natural number, $n > 0$. If a dynamic invariant detector (e.g.\ \Daikon~\cite{daikon07}) is used, 
the following set of invariants is observed:
\begin{itemize}
    \item In the first program, at each iteration of the while-loop: $n > 0$; $sum \ge 0$; $0 \le i \le n$. 
    \item In the second program, at each iteration of the for-loop: $n > 0; s \ge 0; 0 \le j \le n$.
\end{itemize}
Therefore, after renaming some variables ($sum \to s; i \to j$), these two sets of invariants would be considered equivalent. Hence, using sets of invariants allows finding semantically equivalent programs that can differ in their syntax and/or data flow.

Hence, this paper aims to improve the semantic representation of programming exercises using their sets of invariants. These invariants are dynamically detected by \Daikon~\cite{daikon07}, at the beginning and at the end of each scope, over several program executions using a predefined set of test cases for each programming assignment.  In addition to the set of invariants, which provides semantic information about a program, we also leverage the information of a program's structure to its anonymized abstract syntax tree (\AAST), i.e., an \AST after removing all the variables' names.

\paragraph{\textbf{\Clara.}}
\label{sec:clara}
Furthermore, in this paper, we compare our clustering approach against \Clara's clustering method since, to the best of our knowledge, \Clara~\cite{clara} stands as the sole publicly accessible state-of-the-art clustering-based repair tool for repairing \IPAs. 
\Clara leverages correct solutions from past years' submissions for a programming assignment to suggest potential semantic repairs for an erroneous program submitted by a new student.

\paragraph{\Clara's Clustering Approach}
\Clara assigns two programs to the same cluster if they share identical control flow structures and possess a bijective mapping between their variables~\cite{its22-improving-clara-matching-procedure}. However, if there is any deviation in the control flow graphs or a lack of bijective relation between variables, \Clara returns a 'structural mismatch error', resulting in these programs not being clustered together. For each cluster generated, \Clara maintains a \texttt{json} file containing all relevant information about the programs, including expressions and variables.

Revisiting the programs illustrated in Example~\ref{example-cycles}, \Clara does not detect them as matching. Furthermore, when clustering both programs using \Clara's method, they are assigned to separate clusters.

\paragraph{\Clara's Repairing Process.}
To repair an incorrect program, \Clara receives either a single or multiple correct programs, which can be representative of clusters produced by \Clara itself. In such cases, \Clara also necessitates access to all pertinent information regarding the programs contained within each cluster, stored in individual \texttt{json} files. \Clara proceeds to generate a series of repairs for each cluster autonomously, wherein a repair entails adjusting a program to align an expression from the incorrect submission with an expression from a program within the cluster. It is important to emphasize that \Clara's repair suggestions are confined to modifying program expressions and do not involve altering control flow. If \Clara encounters an incorrect program whose control flow does not precisely match any of the correct programs provided, it flags a "Structural Mismatch" error, indicating an inability to repair the program. 
Moreover, \Clara is unable to use the programs outlined in Example~\ref{example-cycles} for repairs, as it does not identify them as matching.

In this paper, our focus is not on enhancing \Clara's repair procedure. Instead, we aim to compare our clustering approach against \Clara's clustering method, and evaluate \Clara's repair performance when utilizing its own clusters versus \InvAAST's clusters of programs.

\section{Syntax Trees and Invariants}
\label{sec:trees-and-invariants}

This section provides some background on syntax trees and program invariants that will be used throughout this paper.

\subsection{Definitions}
\label{sec:preliminaries}

This section provides some definitions that will be used throughout this paper.

\begin{definition}[\textbf{Context-free Grammar (CFG)}]
\label{def:CFG}
A \emph{context-free grammar} $\mathfrak{S}$ is a 4-tuple $(V, \Sigma, R, S)$, where $V$ is the set of non-terminals symbols, $\Sigma$ is the set of terminal symbols, $R$ is the set of rules and $S$ is the start symbol. A CFG describes all the strings permitted in a certain formal language~\cite{hopcroft2008introduction}.
\end{definition}

\begin{definition}[\textbf{Domain-Specific Language (DSL)}]
A Domain-specific Language (DSL) is a tuple $(\mathfrak{S}$, Ops), where $\mathfrak{S}$ is a context-free grammar ($\mathfrak{S}=(V, \Sigma, R, S)$) and Ops is the semantics of DSL operators. The CFG $\mathfrak{S}$ has the rules to generate all the programs in the DSL. The semantics of DSL operators is necessary to analyze conflicts and make deductions.
\end{definition}

\begin{figure}[t!]
    \begin{subfigure}[t!]{0.49\textwidth}
         \centering
          \scalebox{1}{\begin{tikzpicture}[
              level distance=10
              mm,
              level 1/.style={sibling distance=6em},
              every node/.style = {shape=circle,
              draw, align=center,
              top color=white, bottom color=white!20}]]
              \node[label={[label distance=0cm]0:}][draw] at (-5, 0) {decl}
              child { node[label={[label distance=0cm]0:id}] {$i$}}
              child { node[label={[label distance=0cm]0:type}] {$int$}};
          \end{tikzpicture}}
          \caption{AST representation.}
          \label{fig:AST}
     \end{subfigure}
    \begin{subfigure}[t!]{0.49\textwidth}
          \centering
          \scalebox{1}{\begin{tikzpicture}[
              level distance=10
              mm,
              level 1/.style={sibling distance=6em},
              every node/.style = {shape=circle,
              draw, align=center,
              top color=white, bottom color=white!20}]]
              \node[label={[label distance=0cm]0:}][draw] at (-5, 0) {decl}
              child { 
                 node[label={[label distance=0cm]0:id}] {$ID$}
              }
              child { node[label={[label distance=0cm]int:type}] {$int$}};
          \end{tikzpicture}}
          \caption{AAST representation.}
          \label{fig:AAST}
     \end{subfigure}
    \caption{A small example of an AST and an AAST for the variable declaration, \texttt{int i}. An integer variable with identifier $i$.}
    \label{fig:AST-AAST}
\end{figure}

\begin{definition}[\textbf{Abstract Syntax Tree (AST)}]
\label{def:AST}
An \emph{abstract syntax tree (AST)} is a syntax tree in which each node represents an operation, and the children of the node represent the arguments of the operation for a given programming language described by a Context-free Grammar~\cite{hopcroft2008introduction}.
An AST depicts a program's grammatical structure~\cite{compilers-book-dragon}.
\end{definition}

Figure~\ref{fig:AST} presents a small example of the AST representation for the variable declaration \texttt{int i}.

\begin{definition}[\textbf{Anonymized Abstract Syntax Tree (AAST)}]
\label{def:AAST}
An \emph{anonymized abstract syntax tree (AAST)} is an AST in which nodes that contain identifiers are anonymized, i.e., a node's identifier (name of a function or variable) is replaced by a special token (\textit{ID}).
\end{definition}

Figure~\ref{fig:AAST} shows the AAST representation for the same declaration presented previously,~\texttt{int i}.

\begin{definition}[\textbf{Program Invariant}]
\label{def:prog-invariant}
A program invariant is a logical condition that remains true throughout the execution of a program, regardless of its state or inputs. It serves as a guarantee or assertion about the behavior of the program at specific points or during certain operations~\cite{DBLP:books/ph/Dijkstra76}. 
Formally, a program invariant can be defined as a predicate $P$ over the program state variables such that for all program states $s$ reachable during the execution of the program, $P(s)$ holds true.
\end{definition}

Consider again, the first program in Example~\ref{example-cycles}, $n > 0$, $sum \ge 0$, and $0 \le i \le n$ are program invariants that are always satisfied during the execution of the while-loop.

\begin{definition}[\textbf{Bag of Words (BoW)}]
\label{def:BoW}
A \emph{Bag of Words (BoW)} representation~\cite{bag-of-words} is a vector representation where a tokenized sentence is represented as a bag of its words in a vector. The vector representation contains information on the number of times each token in the language appears in the sentence. Note that this model does not take into consideration the language's grammar and even word order. The tokenization step divides a string into $n$-grams,  which are sub-sequences of the original string of $n$ items.
\end{definition}

The following example presents a small illustration of a vector representation of a phrase using a BoW model.

\begin{example}
Let $B$ be a bag of words model computed using the following sentences:
$\{'a a',\- 'e\ i',\- 'a\ e\ i\ o\ u',\- 'o\ i'\}$
Given the phrase $p='a\ i\ a\ u'$, the vector representation of $p$ is, $B(p)\-\ =\-\ [0.5,\-\ 0.0,\-\ 0.25,\-\ 0.0,\-\ 0.25]$. The size of $B(p)$ is 5 since 5 is the size of the vocabulary of $B$. For each entry $s$ of $B(p)$, $B(p)[s]$ corresponds to the percentage of $p$ that is equal to $s$. For example, the symbol $a$ appears twice in a four-symbol phrase. Hence $B(p)[a]=0.5$.
\end{example}

\subsection{Program Invariants}
\label{sec:invariants}
Program invariants are conditions that must always be true at a given point during a program’s execution. Dynamically generated program invariants are \emph{likely invariants} observed during several program executions for a given program. The dynamically generated set of program invariants provides information about a program’s behavior, i.e., its semantics. If two programs share the same program invariants, they are likely semantically equivalent. 
Hence, an invariant-based representation of programs should allow us to find out which
student submissions in a given programming assignment have the same or similar behavior.

\label{sec:var-renaming}
In order to compare two sets of program invariants, a relation between the variables in both sets is required.
We propose to rename all the variables in a program based on the variables’ type and usage. All the variables are renamed the first time they are assigned to some value in a program. The variable’s new name is a concatenation between its type and a counter for how many variables have already been renamed in the program.
With this technique of variable renaming, two programs’ sets of invariants can be easily compared. This method is very simple and fragile, although \IPAs are usually relatively small and simple imperative programs. Therefore, this naive approach should work for \IPAs. Example~\ref{eg:renaming-vars} shows two programs whose variables were renamed using this variable renaming method.

\begin{example}
\label{eg:renaming-vars}
Consider again the programs presented in Example~\ref{example-cycles}, it is important to note that all variables are renamed based on their usage. Specifically, each variable is renamed the first time it is assigned a value in the program. For instance, in the first program, \texttt{n} is renamed to $int_2$ since it is the second variable to be used, in the \texttt{scanf} (line 2). After renaming all variables based on their usage, the following mapping of variables for the first program is obtained: \{$sum \to int_0$; $n \to int_1$; $i \to int_2$\}. Applying the same procedure to the second program yields the mapping \{$s \to int_0$; $n \to int_1$; $j \to int_2$\}. Thus, the two programs after renaming are as follows:
          
\begin{minipage}[t!]{0.48\linewidth}
\begin{code}
\begin{minted}[escapeinside=çç,tabsize=1,obeytabs,xleftmargin=5pt,linenos]{C}
int int1, int0 = 0, int2;
scanf("%d", &int1);
int2 = 0;
while(int2 < int1){
  int2++;
  int0 = int0 + int2;
}
printf("%d\n",int0);
\end{minted}
\end{code}
\end{minipage}
\begin{minipage}[t!]{0.48\linewidth}
\begin{code}
\begin{minted}[escapeinside=çç,tabsize=1,obeytabs,xleftmargin=5pt,linenos]{C}
int int2,int1,int0 = 0;
scanf("%d", &int1);

for(int2 = int1; int2 >= 0; int2--)
{
  int0 = int2 + int0;
}
printf("%d\n", int0);
\end{minted}
\end{code}
\end{minipage}
\vspace{0.5cm}

Hence, the set of invariants of both cycles (for and while) is the same: \{$int_1>0$; $int_0 \ge 0$;  $0 \le int_2 \le int_1$\}.
\end{example}

In this work, we use \Daikon~\cite{daikon07} to compute dynamically-generated likely invariants observed across multiple program executions for each student submission, employing a predefined set of input-output tests for each programming assignment. \Daikon's default format for program invariants combines Java and mathematical logic, aiming to convey meaning concisely to programmers.

To adapt \Daikon for small imperative C programs, we initially apply a method for variable renaming to all student submissions. Next, we inject empty functions into each scope and pass the variables of the respective scope as parameters. A scope is defined as each block of statements without branching, and in cases of nested scopes, we include all variables available in the parent scopes as parameters as well.
We adhere to conventional methods of modeling control flows, such as, computing invariants before a loop, before a loop guard, inside a loop guard, inside the loop, and after the loop.

Finally, \Daikon is executed using all input tests for each programming assignment. The dynamically-generated invariants produced by \Daikon  are stored for each program's structure or scope (e.g., \texttt{if statements, loops, blocks}).
We do not specifically ask \Daikon to generate any type of invariant. The only type of invariants we turned off is the “OneOf” invariants (e.g., “x is OneOf \{1,2\}”) that may cause overfitting to the test suite. Example~\ref{eg:injected_progs} presents the two programs, previously presented in Example~\ref{eg:renaming-vars}, after being injected with empty functions in each program scope.

\begin{example}
\label{eg:injected_progs}
For Daikon~\cite{daikon07} to work on small imperative programs, we have to call empty functions at the beginning of each scope and pass all the visible variables, in that scope, as parameters. 

Consider again the programs presented in Example~\ref{eg:renaming-vars}, after renaming all the variables based on their usage. These programs, after being injected with empty functions, would look like the following programs:

\begin{minipage}[t!]{0.48\linewidth}
\begin{code}
\begin{minted}[escapeinside=çç,tabsize=1,obeytabs,xleftmargin=5pt,linenos]{C}
scope_1();                  
int int1, int0 = 0, int2;
scanf("%d", &int1);
int2 = 0;
while(int2 < int1, 
while_1(int0, int1, int2)){
  scope_2(int0, int1, int2);
  int2++;
  int0 = int0 + int2;
}
scope_3(int0, int1, int2);
printf("%d\n",int0);
\end{minted}
\end{code}
\end{minipage}
\begin{minipage}[t!]{0.48\linewidth}
\begin{code}
\begin{minted}[escapeinside=çç,tabsize=1,obeytabs,xleftmargin=5pt,linenos]{C}
scope_1();                       
int int2, int1, int0 = 0;
scanf("%d", &int1);

for(int2 = int1; int2 >= 0, 
for_1(int0, int1, int2); int2--)
{
  scope_2(int0, int1, int2);
  int0 = int2 + int0;
}
scope_3(int0, int1, int2);
printf("%d\n", int0);
\end{minted}
\end{code}
\end{minipage}
\vspace{0.5cm}

\end{example}

\section{Program Representations}
\label{sec:representations}

As the primary objective of this work is to cluster programs based on semantic and syntactic features, each program is represented as a feature vector.
In particular, we propose to use a \emph{bag of words (BoW)} model (see Definition~\ref{def:BoW}).
Using BoW models, we generate vector representations for each student submission based on several features. These features may include the Abstract Syntax Tree (AST), set of invariants, or even the program code. It is also possible to combine several of these features. Next, all the vector representations used in this work are described.

\subsection{Syntax Vectors}
\label{sec:syntax-vector}
The syntax vector program representation is the simplest to compute since it is based solely on the program syntax (code). In the interest of comparing the syntax of programs independently of the variables' names, first, all the programming solutions are renamed using the method described in Section~\ref{sec:var-renaming}. Next, all the student submissions are tokenized, and a vocabulary with all the available tokens is obtained. Then, vectors for each student submission are created, where the $i^{th}$ entry is the number of times the $i^{th}$ word of the vocabulary appears in the program. Finally, the numbers of occurrences in these vectors are normalized.

\subsection{Anonymized Abstract Syntax Tree Vectors}
\label{sec:aast}
An alternative representation is to compute a bag of words using the strings of the abstract syntax trees (ASTs) of all submissions for a given programming assignment. 
This representation has already been used in program clustering~\cite{sarfgen, deckard}. 
However, we represent each AST as a string and remove all names of variables and functions, keeping only their respective types in the AST.
Thus, for each submission, we have an \emph{anonymized abstract 
syntax tree (\AAST)} (see Definition~\ref{def:AAST}). With these \AASTs, we keep the information about a program's structure, ignoring the name of its variables. The information about a program's structure is kept since an \AAST contains all the non-terminal symbols of the language's grammar. Next, a vocabulary is built with the tokens present in all submissions, and a normalized vector representation for each \AAST is computed.

\subsection{Invariant Vectors}

\label{sec:invariant}
Another approach is to use an invariant-based vector representation. 
In this case, we apply the bag of words model to the set of invariants of the programs. 
We gather all program invariants as described in Section~\ref{sec:invariants}. 
Previous work on the use of invariants to detect semantic similarity between possible patches to a program~\cite{ase19-weimer-understanding-patches-through-invariants} showed that using string distance measure between invariant sets had similar results and was more efficient than computing the logical similarity between their corresponding sets of program invariants. 
Therefore, we represent our invariants in the form of strings. However, instead of using a string distance measure between invariant sets (e.g., \ Levenshtein edit distance~\cite{levenshtein-dist}), we create a bag of words model with those sets of invariants.

\subsection{Combination of Program Features}
\label{sec:features-combinations}
Finally, observe that these vector representations (\textit{Syntax, \AAST, Invariants}) can be combined, thus taking advantage of using
several types of features. 
For example, we use a bag of words in our work using the program's AST and
the sets of invariants. In this case, first, we build two BoW representations independently, one based on AASTs and another one based on invariants. Then, we concatenate, for each submission, the submission's vector representations using the two BoWs, achieving a vector representation based on the program's AAST and set of invariants.
Both vectors, from the Invariants BoW and the \AAST BoW, are normalized before concatenating. Therefore, the invariants and the \AASTs have an equal contribution to the \AAST+ Invariants BoW.
The program syntax was not included in this last representation since the BoW based on syntax has a large vocabulary that generates vectors that are too sparse.

\begin{figure*}
    \centering
    \scalebox{0.45}{
    \includegraphics{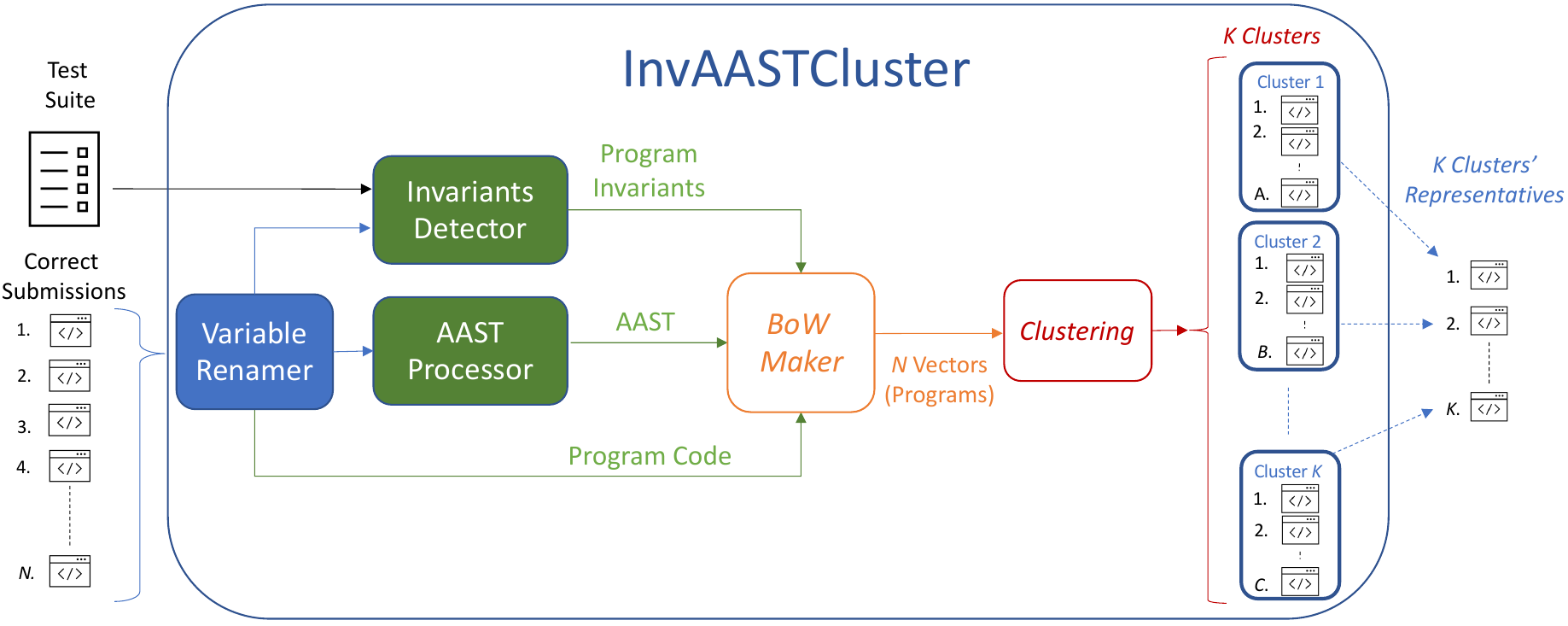}}
    \caption{The high-level overview of \InvAAST.}
    \label{fig:invaast}
\end{figure*}

\section{Implementation}
\label{sec:implementation}

This section presents the implementation of our program clustering technique. We implemented the proposed approach in the tool \InvAAST (\textbf{Inv}ariants and \textbf{AAST} Program \textbf{Cluster}ing). \InvAAST is publicly available on GitHub at \href{https://github.com/pmorvalho/InvAASTCluster}{https://github.com/\-pm\-or\-va\-lho/\-Inv\-AAST\-Clus\-ter}.
Figure~\ref{fig:invaast} shows the overall architecture of \InvAAST. Given a set of $N$ correct submissions and a test suite, \InvAAST computes $K$ clusters of programs ($N \ge K$) and returns the set of $K$ clusters' representatives, i.e., the set of correct programs that are closest to the center of each one of the $K$ clusters. \InvAAST is divided into six main modules: variable renamer, invariants detector, AASTs pro\-ce\-ssor, bag of words (BoW) maker, clustering procedure, and the selection of each cluster representative.

\paragraph{Variable Renamer.} In this module, \InvAAST renames all variables of each one of the $N$ given correct submissions. All variables are renamed based on their usage in each program, as explained in Section~\ref{sec:var-renaming}. \InvAAST uses \texttt{pycparser}\footnote{\href{https://github.com/eliben/pycparser}{https://github.com/eliben/pycparser}} to find all variables in a program. Then, when a variable is first used in the program (e.g. a\-ssign\-ment) that variable receives a new name considering the variable's type. 

\paragraph{Invariants Detector.} \InvAAST uses \Daikon~\cite{daikon07} to compute dy\-na\-mi\-ca\-lly-ge\-ne\-ra\-ted invariants for a given test suite (see Section~\ref{sec:var-renaming}). 
After all the variables have been renamed, this module produces a set of invariants for each program's scope using the provided test suite.
All these sets of invariants are then sent to the BoW maker module.

\paragraph{AAST Processor.} In this step, \InvAAST also uses \texttt{pycparser} to compute a program's abstract syntax tree (AST). Additionally, \InvAAST removes all the variables' and functions' identifiers from the AST to transform the program's AST into an anonymized abstract syntax tree (AAST), conserving only the program's structure.

\paragraph{Bag of Words (BoW) Maker.} This module receives three sets as input: (1) the set of correct program submissions with all their variables renamed from the Variable Renamer module; (2) all the program's AASTs from the AAST processor, and (3) the set of the programs' dy\-na\-mi\-ca\-lly-ge\-ne\-ra\-ted invariants. The \textit{BoW Maker} computes the bag of words (BoW) model (see Definition~\ref{def:BoW}) that is going to be used to generate vector representations for each program. 

Depending on this module's parameterization, the BoW maker can compute four different bags of words: (1) based on the programs' code (syntax), (2) based on the programs' AASTs (structure), (3) using the set of programs' invariants (semantics) and (4) joining the programs' AASTs and their sets of invariants (structure + semantics).
To compute these BoW models, \InvAAST uses scikit-learn package, \texttt{feature\_extraction}\footnote{\href{https://scikit-learn.org/stable/modules/generated/sklearn.feature_extraction.text.TfidfVectorizer.html}{sklearn.feature\_extraction.text.TfidfVectorizer}}. 

\InvAAST tokenizes the input strings into tokens of size $n$ ($n$-grams) to build a vocabulary with all the submissions' information, i.e., invariants, syntax, or AASTs. 
In our case, we define $n = 3$ ($3$-grams) for this parameter of the BoW maker.
Afterward, once a vocabulary has been collected, \InvAAST computes a vector representation for each program by counting the number of times each token appears in the program's information string (invariants, syntax, or AASTs) normalizing the vector by the length of the BoW's vocabulary.

\paragraph{Clustering.} 
The main goal of \InvAAST is to reduce the vast number of correct submissions, $N$, into a significantly smaller number of program representatives, $K$, to help program repair frameworks to become more scalable (if $N \gg K$). Therefore, 
\InvAAST accepts as parameter the number of desired clusters $K$, which is by default $10\%$ of $N$.
The BoW maker module passes the set of vector representations for each one of the $N$ correct submissions to the clustering procedure. Then, \InvAAST uses the \emph{KMeans} algorithm to cluster these submissions into $K$ different clusters.
The \emph{KMeans} algorithm receives as a parameter the number of clusters it should return ($K$). The KMeans algorithm divides the set of observations, in our case, students' programs, into $K$ clusters, where each program is assigned to the cluster with the nearest mean~\cite{k-means}. 
\InvAAST uses KMeans but other clustering algorithms can be applied. 
\InvAAST provides users with the option to select from various clustering algorithms offered in scikit-learn~\footnote{\href{https://scikit-learn.org/stable/modules/clustering.html}{https://scikit-learn.org/stable/modules/clustering.html}}: Affinity Propagation, MeanShift, MiniBatch KMeans, Agglomerative Clustering, Ward, Spectral Clustering, DBSCAN, OPTICS, BIRCH, and Gaussian Mixture.

\paragraph{Clusters' representatives selection.} In this last module, \InvAAST chooses a program representative for each cluster. For each one of the $K$ clusters, \InvAAST computes the program closest to the center of the cluster using the Euclidean distance. Afterward, \InvAAST returns a set of $K$ clusters' representatives.

\begin{figure}[t!]
    \centering
    \scalebox{0.5}{
    \includegraphics{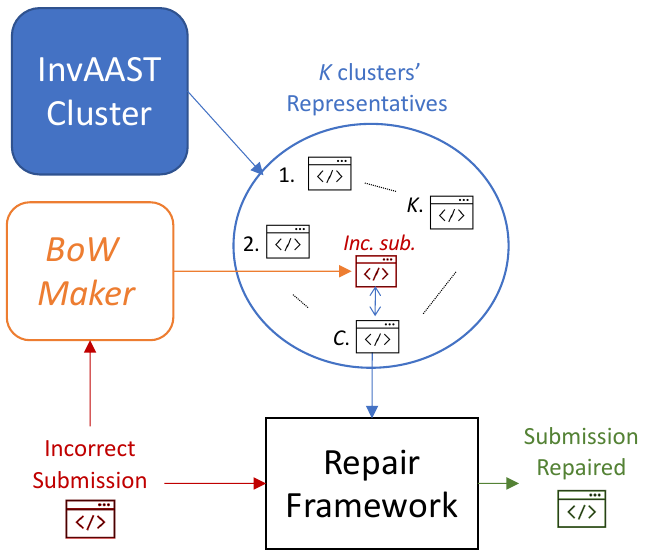}}
    \caption{Finding the closest correct program, i.e., the closest correct program representative to the incorrect submission vector representation. This approach passes only one program to the repair tool instead of $K$ programs.}
    \label{fig:closest-program}
\end{figure}

\mbox{}\hfill

\emph{Easy upgradability.} \InvAAST was designed with modularity in mind. On that account, one can easily remove, add or modify any module of \InvAAST. For example, one can use other models instead of the bag of words model, only needing to replace that specific procedure.

\mbox{}\hfill

Several repair tools~\cite{refactory, autograder, verifix} are implementation-based program repair tools, i.e., they receive a single correct program to act as a reference implementation for repairing any given incorrect program. Hence, these frameworks are not designed to take advantage of a vast number of semantically different correct submissions. These frameworks can be run in parallel. However, they typically do not have a procedure to choose which among several possible repairs is the best (minimal). These tools compute the set of repairs based on the reference implementation.
\InvAAST cannot be used on these frameworks since its output is a set of $K$ clusters' representatives. Consequently, in order to allow these non-clustering-based frameworks to take advantage of \InvAAST, we developed an additional module that finds the closest correct program representative to an incorrect program. Our motivation is that with the closest correct submission from previous years, these tools can provide the student with a minimal set of repairs.

\paragraph{Closest correct program finder.}
\label{sec:implementation-closest-prog}
The overall idea of this module is presented in Figure~\ref{fig:closest-program}. Given a student's incorrect submission, \InvAAST finds which of the $K$ clusters' representatives, returned by \InvAAST's selection module, is the closest program to the incorrect submission. This is done by identifying the smallest Euclidean distance between the vector representation of each one of the clusters' representatives and the incorrect submission. Hence, we can identify one correct program that is most likely the reference implementation to use for repairing a specific student's program. In the example of Figure~\ref{fig:closest-program}, \InvAAST would return only the program $C$  to the repair framework since it is the closest program to the incorrect submission.

\section{Introductory Programming Assignments (\IPAs) Datasets}
\label{sec:datasets-ipas}

\paragraph{\textbf{\benchmark}.} In order to evaluate the program representations described in Section~\ref{sec:representations}, we gathered a benchmark of students' programs developed during an introductory programming course in C language. These programs were collected over three distinct practical classes at Instituto Superior Técnico for 25 different \IPAs. Each lab class focuses on a different topic of the C programming language. Lab02 deals with integers and input-output operations. Secondly, Lab03 focuses on loops and chars. Lastly, in Lab04, the students learn to use vectors and strings.
The textual description of each programming assignment can be found in Appendix~A.

This dataset of introductory programming exercises, \benchmark, is publicly available at ~\href{https://github.com/pmorvalho/C-Pack-IPAs}{https://\-github.com/\-pmorvalho/\-C\--\-Pack-\-IPAs} and can be used by other \IPAs repair/clustering frameworks. In this git repository, the interested reader can find all the information about the description and the input-output tests used to evaluate each \IPA~\cite{C-Pack-IPAs}. Furthermore, there is also a reference implementation for each \IPA in the public git repository that can be used by program repair frameworks that only accept a single reference implementation to repair incorrect programs. Moreover, \benchmark~\cite{C-Pack-IPAs_apr24} has also proven successful in evaluating various works across program analysis~\cite{aitp22-learning-2-map-variables,ecai23-GNNs-4-var-mapping}, fault localization~\cite{CFaults-fm24}, program repair~\cite{DBLP:conf/aaai/OrvalhoJM25} and program transformation~\cite{fse22-MultIPAS}.

Since this work focuses only on program semantics, only submissions that compile without any errors were selected.
The set of submissions was split into two sets: correct submissions and incorrect submissions. 
The students' submissions that satisfied a set of input-output test cases for each \IPA were considered correct and selected as benchmark instances. The submissions that failed at least one input-output test were considered incorrect. 

Table~\ref{tab:table-IPAs} presents the number of submissions gathered. For 25 different programming exercises, this dataset contains 1620 different correct and 196 incorrect submissions.
\Clara's clustering method does not support all the features present in the correct submissions collected. Hence, as shown in Table~\ref{tab:table-IPAs}, after removing the set of exercises and correct programs that \Clara does not support, we achieved a final set of 1141 correct submissions for 20 \IPAs.

\paragraph{\textbf{ITSP}.} The \textsc{ITSP} dataset has been used by other automated program analysis tools~\cite{asr-for-ITSP, verifix, fse22-MultIPAS}. 
This dataset is also a collection of C programs although it is well balanced, i.e., the number of correct submissions is closer to the number of incorrect submissions in this dataset. Table~\ref{tab:itsp-dataset} presents the number of programs in the \textsc{ITSP} dataset after we removed the programs that \Clara and our variable renamer module do not support.

\begin{table}[t!]
\centering
\caption{Description of our dataset of \IPAs.
}
\label{tab:table-IPAs}
\scalebox{1}{\begin{tabular}{|c|c|c|c|c|c|}
\hline
\textbf{Labs} &
  \textbf{\#IPAs} &
  \textbf{\begin{tabular}[c]{@{}c@{}}\#Correct \\ Submissions\end{tabular}} &
  \textbf{\begin{tabular}[c]{@{}c@{}}\#Incorrect \\ Submissions\end{tabular}} &
  \textbf{\begin{tabular}[c]{@{}c@{}}\#IPAs\\ (\Clara)\end{tabular}} &
  \textbf{\begin{tabular}[c]{@{}c@{}}\#Correct \\ Submissions\\ (\Clara)\end{tabular}} \\ \hline
\textbf{Lab02} & 10          & 789           & 118          & 10          & 738           \\ \hline
\textbf{Lab03} & 7           & 363           & 35           & 5           & 244           \\ \hline
\textbf{Lab04} & 8           & 468           & 43           & 5           & 159           \\ \hline
\textbf{Total} & \textbf{25} & \textbf{1620} & \textbf{196} & \textbf{20} & \textbf{1141} \\ \hline
\end{tabular}}
\end{table}

\begin{table}[t!]
\centering
\caption{Description of ITSP~\cite{asr-for-ITSP} dataset. Correct programs that our approach and \Clara do not support were removed.}
\label{tab:itsp-dataset}
\scalebox{1}{\begin{tabular}{|c|c|c|c|}
\hline
\textbf{\begin{tabular}[c]{@{}c@{}}ITSP\\ Dataset\end{tabular}} &
  \textbf{\#IPAs} &
  \textbf{\begin{tabular}[c]{@{}c@{}}\#Correct \\ Submissions\end{tabular}} &
  \textbf{\begin{tabular}[c]{@{}c@{}}\#Incorrect \\ Submissions\end{tabular}} \\ \hline
\textbf{Lab3}  & 4           & 45           & 63           \\ \hline
\textbf{Lab4}  & 6           & 74           & 75           \\ \hline
\textbf{Lab5}  & 7           & 64           & 62           \\ \hline
\textbf{Lab6}  & 6           & 19           & 24           \\ \hline
\textbf{Total} & \textbf{23} & \textbf{202} & \textbf{224} \\ \hline
\end{tabular}}
\end{table}

\section{Experiments}
\label{sec:results}
The experimental results presented in this section aim to support our claims that the proposed novel program representation based on a program's \AAST and its set of program invariants help (1) to efficiently cluster semantically equivalent small imperative programs submitted in \IPAs, and (2) to repair faster and significantly more \IPAs' incorrect submissions in current state-of-the-art clustering-based program repair tools, such as \Clara~\cite{clara}.

The goal of our experiments was to answer the following research questions:\\
\textbf{RQ1.} How does invariant-based program clustering compare against \AAST and syntax-based clustering on a set of correct submissions? 
(Section~\ref{sec:results-clustering})
\\
\textbf{RQ2.} Does \Clara repair more programs using \InvAAST's closest correct submission or its set of \textsc{KMeans} clusters' representatives?~(Section~\ref{sec:results-repair})

To answer these research questions, we evaluate \InvAAST in two different use cases: (1)~clustering \IPAs~(Section~\ref{sec:results-clustering}), and (2)~repairing \IPAs~(Section~\ref{sec:results-repair}). For this evaluation, we have gathered a set of \IPAs, previously described in Section~\ref{sec:datasets-ipas}, developed during an introductory programming university course in C language. Section~\ref{sec:results-clustering} presents the first use case where we perform clustering on the students' submissions for different \IPAs and evaluate its accuracy on different 
program representations. Afterward, Section~\ref{sec:results-repair} shows the  second use case where we integrate our program representations into a state-of-the-art program repair tool, \Clara~\cite{clara}, to evaluate if our clustering technique is able to outperform \Clara's clustering method, which is the only current publicly available state-of-the-art clustering method for repairing \IPAs. 
We are going to give our program clusters for each \IPA to \Clara and use \Clara's repairing process. The idea is to evaluate our clustering approach being integrated into a clustering-based program repair tool. Program repair is only one of the several possible applications for our program clustering method.

All of the experiments were conducted on an Intel(R) Xeon(R) Silver computer with
4210R CPUs @ 2.40GHz, using a memory limit of 64GB.

\subsection{Use Case 1: Clustering \IPAs}
\label{sec:results-clustering}

\begin{table}[t!]
\centering
\caption{The values for the cluster accuracy using four different clustering algorithms on each program representation after ten different runs, each run using a different seed.}
\label{tab:cluster-acc}
\resizebox{\columnwidth}{!}{%
\begin{tabular}{|c|l|c|c|c|c|}
\hline
\textbf{Clustering Algorithm} &
  \multicolumn{1}{c|}{\textbf{Program Representation}} &
  \textbf{Average} &
  \textbf{Median} &
  \textbf{Variance} &
  \textbf{Standard deviation} \\ \hline
\multirow{4}{*}{\textbf{\textsc{KMeans}}} &
  AAST+Invariants &
  \textbf{81.44\%} &
  \textbf{80.65\%} &
  0.02\% &
  1.35\% \\ \cline{2-6} 
                                           & AAST            & 73.63\%          & \textbf{73.70\%} & 0.03\% & 1.69\% \\ \cline{2-6} 
                                           & Invariants      & \textbf{78.69\%} & \textbf{78.73\%} & 0.01\% & 0.88\% \\ \cline{2-6} 
                                           & Syntax          & 58.05\%          & 58.15\%          & 0.01\% & 1.22\% \\ \hline
\multirow{4}{*}{\textbf{MiniBatch \textsc{KMeans}}} & AAST+Invariants & 79.09\%          & 79.63\%          & 0.05\% & 2.23\% \\ \cline{2-6} 
                                           & AAST            & 72.33\%          & 72.96\%          & 0.10\% & 3.12\% \\ \cline{2-6} 
                                           & Invariants      & 75.83\%          & 76.23\%          & 0.03\% & 1.68\% \\ \cline{2-6} 
                                           & Syntax          & 58.46\%          & 58.21\%          & 0.03\% & 1.81\% \\ \hline
\multirow{4}{*}{\textbf{Birch}}            & AAST+Invariants & 80.10\%          & 80.09\%          & 0.00\% & 0.51\% \\ \cline{2-6} 
                                           & AAST            & \textbf{73.92\%} & 73.55\%          & 0.08\% & 2.81\% \\ \cline{2-6} 
                                           & Invariants      & 77.99\%          & 77.90\%          & 0.01\% & 0.86\% \\ \cline{2-6} 
                                           & Syntax          & \textbf{59.05\%} & \textbf{58.98\%} & 0.01\% & 0.94\% \\ \hline
\multirow{4}{*}{\textbf{Gaussian Mixture}} & AAST+Invariants & 78.89\%          & 79.60\%          & 0.05\% & 2.26\% \\ \cline{2-6} 
                                           & AAST            & 70.97\%          & 71.70\%          & 0.09\% & 2.92\% \\ \cline{2-6} 
                                           & Invariants      & 75.59\%          & 74.81\%          & 0.07\% & 2.63\% \\ \cline{2-6} 
                                           & Syntax          & 57.70\%          & 57.35\%          & 0.02\% & 1.58\% \\ \hline
\end{tabular}%
}
\end{table}

\begin{figure*}[t!]
    \begin{subfigure}[t!]{0.49\textwidth}
         \includegraphics[width=\textwidth]{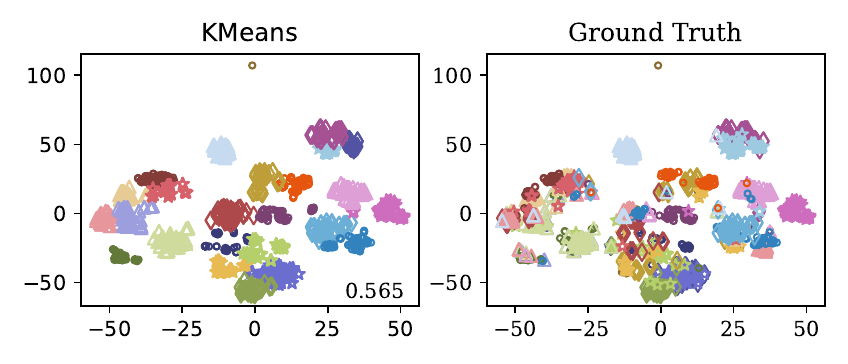}
         \caption{Syntax  Representation}
         \label{fig:syntax}
     \end{subfigure}
     \hfill
    \begin{subfigure}[t!]{0.49\textwidth}
         \includegraphics[width=\textwidth]{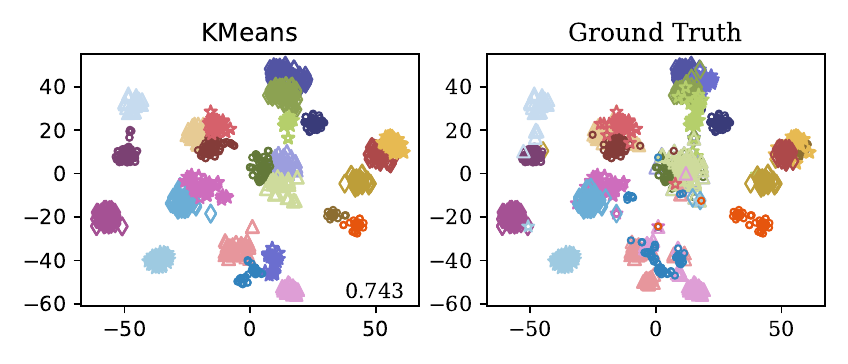}
         \caption{AAST Representation}
         \label{fig:aast}
     \end{subfigure}
     \begin{subfigure}[t!]{0.49\textwidth}
        \includegraphics[width=\textwidth]{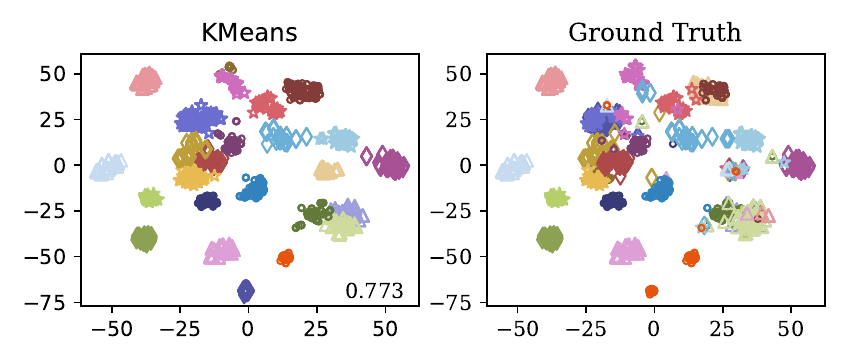}      \caption{Invariant-Based Representation}
        \label{fig:invariants}
     \end{subfigure}
     \hfill
    \begin{subfigure}[t!]{0.49\textwidth}
         \includegraphics[width=\textwidth]{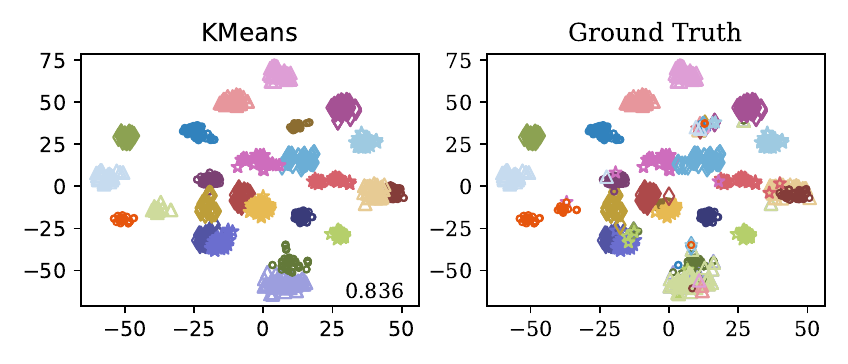}
         \caption{AAST + Invariants Representation}
         \label{fig:aast-invariants}
     \end{subfigure}
    \caption{Comparison between the ground truth (on the right) and the clusters and cluster accuracy obtained using the \textsc{KMeans} algorithm (on the left) for each type of program representation.}
    \label{fig:clustering-plots}
\end{figure*}

A study was performed to evaluate different program representations by applying program clustering to the set of correct programs described in Table~\ref{tab:table-IPAs}. 
The main idea of this experiment was to evaluate if program invariants help identify different \IPAs' submissions. 

\InvAAST was used to cluster the 1620 correct submissions, to different \IPAs, into 25 distinct clusters since our dataset has 25 different programming exercises.
Other works~\cite{semCluster-pldi19} that perform program clustering on \IPAs perform an equivalent study on clustering submissions for different exercises.
The main reason to cluster programs from 25 different exercises is that we know the ground truth label for each program since we know for which specific \IPA the students submit their assignments. Otherwise, we would have to manually choose semantically different implementations for the same \IPA and assign labels, which might be subjective.

\InvAAST, as explained in Section~\ref{sec:implementation}, starts by renaming all the variables in the student submissions. Then uses \Daikon~\cite{daikon07} to collect the student submissions' dynamically generated invariants sets as described in Section~\ref{sec:invariants}. Lastly, it uses the python library, \texttt{pycparser}\footnote{\href{https://github.com/eliben/pycparser}{https://github.com/eliben/pycparser}}, to compute all the anonymized abstract syntax trees (AAST) (see Section~\ref{sec:aast}). Using all these program features, we computed four different bags of words models. One model for each program representation (syntax, AAST, and invariants) and one additional model using a combination of a program's AAST and its invariants set. The program syntax is not included in this last representation since the bag of words based on program syntax has a large vocabulary that generates vectors that are too sparse.

The following clustering algorithms available in scikit-learn\footnote{\href{https://scikit-learn.org/stable/modules/clustering.html}{https://scikit-learn.org/stable/modules/clustering.html}} were applied to each program representation: \textsc{KMeans}, MiniBatch \textsc{KMeans}, BIRCH and Gaussian Mixture. 
The \emph{KMeans} algorithm divides the set of observations, in our case students' programs, into $n$ clusters where each program is assigned to the cluster with the nearest mean. We used the Euclidean distance as the similarity measurement for \textsc{KMeans}.
MiniBatch \textsc{KMeans} is similar to KMeans, but instead of using the entire dataset to update the cluster centers at each iteration, this algorithm uses randomly selected subsets. On the other hand, BIRCH is a hierarchical clustering algorithm that builds a tree structure to represent the data. It incrementally clusters data points by recursively splitting clusters into subclusters. Finally, a Gaussian Mixture Model is a weighted sum of $n$ component Gaussian densities where $n$ is the number of clusters~\cite{gaussian-mixture}. Our discussion centers around these clustering algorithms, as they yielded the most favorable outcomes in our experiments.

Since our dataset of \IPAs has 25 different programming exercises, the ground truth has 25 different clusters.
Each student program is a submission to a specific programming exercise (label) that we know. Consequently, the \emph{cluster accuracy} metric can be used to evaluate the obtained clusters. With this metric, each cluster is assigned the label (exercise) which is most frequent in the cluster. Afterward, the accuracy of this assignment is measured by counting the number of correctly assigned student submissions and dividing by the number of total submissions. This metric is also known as \emph{purity}~\cite{cluster-purity}.

Table~\ref{tab:cluster-acc} presents the average, median, variance, and standard deviation values for the cluster accuracy for each clustering algorithm on four different program representations after ten different runs. Each run uses a different seed. Entries highlighted in bold correspond to the highest average/median accuracy values for each different program representation for all clustering algorithms.
One can see that the \AAST+ Invariants representation has the best performance considering all the clustering algorithms. The Invariants BoW has the second highest accuracy, followed by the \AAST BoW. Lastly, the Syntax BoW presents the poorest performance of all. 
From now on, we focus the discussion on the \textsc{KMeans} results since this clustering algorithm achieved the best results (see Table~\ref{tab:cluster-acc}). 

\begin{figure}[t!]
    \centering
    \scalebox{1}{
    \includegraphics{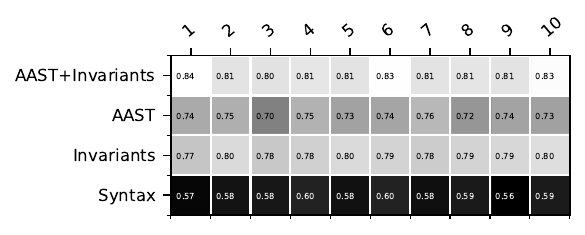}
    }
    \caption{The values for cluster accuracy using the \textsc{KMeans} algorithm on each program representation after ten different runs, each run using a different seed.}
    \label{fig:acc-kmeans}
\end{figure}

The \textsc{KMeans} clustering algorithm divides the set of observations, in our case, students' programs, into $n$ clusters where each program is assigned to the cluster with the nearest mean~\cite{k-means}.
The \textsc{KMeans} algorithm receives as a parameter the number of clusters it should return, i.e., it always returns 25 different clusters of programs. 
Figure~\ref{fig:acc-kmeans} shows a matrix with the different cluster accuracy values using the \textsc{KMeans} algorithm on each program representation for ten different seeds. Each entry is highlighted accordingly to its value. The lowest value is highlighted in black, and the highest is highlighted in white. Intermediate values are highlighted in different shades of grey, depending on how far they are from the lowest value. Matrices with values of the cluster accuracy for the other clustering algorithms can be found in Appendix~\ref{appendix:clustering-acc}.

Furthermore, Figure~\ref{fig:clustering-plots} presents the results of applying the \textsc{KMeans} model to each one of the four program representations being analyzed. 
To present these results graphically, we used a method for visualizing high-dimensional data in a $2-$dimension map, called \textsc{t-SNE}~\cite{tsne}.
Each subfigure corresponds to a different type of representation. The left side of each subfigure shows the clustering results and the value of the cluster accuracy (right-bottom corner). The right side presents the real clusters of each programming exercise, i.e., the ground truth represented using each program representation.
Figure~\ref{fig:syntax} shows the results after clustering all the student submissions using a syntax representation, which resulted in a cluster accuracy of almost $57\%$. The AAST representation achieved a cluster accuracy of $74.3\%$ as presented in Figure~\ref{fig:aast}. 

Regarding the use of program invariants, Figure~\ref{fig:invariants} and Table~\ref{tab:cluster-acc} support the idea that program invariants improve program clustering since this representation obtained a cluster accuracy of $77.3\%$. Lastly, Figure~\ref{fig:aast-invariants} presents the representation that uses the combination AASTs and invariants sets, which also shows an improvement compared to the invariant-based representation. This representation outperforms all the other representations with an accuracy of $83.6\%$. Furthermore, Table~\ref{tab:cluster-acc} shows that this representation based on AAST and invariants achieved the best cluster accuracy for all clustering algorithms.
Another advantage of this representation is that it best separates all the students' submissions in different regions of the space, i.e., the majority of the clusters are visibly separated from each other.

Other evaluation metrics for the \textsc{KMeans} algorithm can be found in Appendix~\ref{appendix:clustering-metrics} for the interested reader's convenience. Clustering metrics such as the \emph{Rand index}, the \emph{adjusted Rand index}, the \emph{normalized mutual information}, the \emph{adjusted mutual information}, the \emph{Fowlkes–Mallows index}, the \emph{completeness score}, the \emph{homogeneity score}, and the \emph{V measure}. 

\subsection{Use Case 2: Repairing \IPAs} 
\label{sec:results-repair}

This section presents the results of integrating \InvAAST as the clustering approach for \Clara~\cite{clara}, a publicly available state-of-the-art clustering-based program repair tool.
Since our set of \IPAs, described in Table~\ref{tab:table-IPAs}, has a small number of incorrect submissions, only 196, for this evaluation, we have also considered the \textsc{ITSP} dataset~\cite{asr-for-ITSP} described in Section~\ref{sec:datasets-ipas}. 
Thus, overall we have a total of 420 incorrect submissions (196 from our dataset plus 224 from the \textsc{ITSP} dataset) and 1343 correct submissions (1141 from our dataset plus 202 from the \textsc{ITSP} dataset) for 43 different \IPAs (20 from our dataset plus 23 from the \textsc{ITSP} dataset). To fully evaluate our clustering technique for repairing \IPAs, we are going to compare \InvAAST's results against \Clara's in terms of: (1) the number of student submissions repaired; (2) the number of clusters produced by each clustering approach for each \IPA; 
and (3) the time spent to repair each incorrect submission.

We would like to point out that in this experiment, we are not trying to improve \Clara's repair process. Instead, we are comparing the performance of \Clara's repair process when using its own or \InvAAST's clusters of programs.

\begin{table}[t!]
\centering
\caption{This table presents the percentage of submissions repaired (success), structural mismatch errors, and timeouts (failure) for each clustering approach. The total number of submissions is 319.}
\label{tab:repair-results}
\scalebox{1}{
\begin{tabular}{cl|c|cc|}
\cline{3-5}
\multicolumn{1}{l}{}             &                      & \textbf{\%Success} & \multicolumn{2}{c|}{\textbf{\%Failure}}                  \\ \cline{2-5} 
\multicolumn{1}{l|}{} &
  \textbf{Clustering Method} &
  \textbf{\begin{tabular}[c]{@{}c@{}}\%Submissions \\ Repaired\end{tabular}} &
  \multicolumn{1}{c|}{\textbf{\begin{tabular}[c]{@{}c@{}}\%Structural\\ Mismatch\end{tabular}}} &
  \textbf{\begin{tabular}[c]{@{}c@{}}\%Timeouts\\ (600s)\end{tabular}} \\ \hline
\multicolumn{1}{|c|}{\textbf{1}} & \Clara                & 71.79\%            & \multicolumn{1}{c|}{7.52\%}           & \textbf{20.69\%} \\ \hline
\multicolumn{1}{|c|}{\textbf{2}} & \textsc{KMeans} - Invs        & 82.45\%            & \multicolumn{1}{c|}{\textbf{11.91\%}} & 5.64\%           \\ \hline
\multicolumn{1}{|c|}{\textbf{3}} & \textsc{KMeans} - Syntax      & 84.01\%            & \multicolumn{1}{c|}{10.97\%}          & 5.02\%           \\ \hline
\multicolumn{1}{|c|}{\textbf{4}} & \textsc{KMeans} - AAST        & 84.64\%            & \multicolumn{1}{c|}{9.72\%}           & 5.64\%           \\ \hline
\multicolumn{1}{|c|}{\textbf{5}} & \textsc{KMeans} - AAST + Invs & \textbf{84.95\%}   & \multicolumn{1}{c|}{10.03\%}          & 5.02\%           \\ \hline
\multicolumn{1}{|c|}{\textbf{6}} &
  \begin{tabular}[c]{@{}l@{}}Closest Program (\textsc{KMeans}) \\ - AAST + Invs\end{tabular} &
  84.33\% &
  \multicolumn{1}{c|}{10.66\%} &
  5.02\% \\ \hline
\end{tabular}%
}
\end{table}

\label{sec:results_repair_subsection} 
Different procedures for program clustering using \InvAAST (see Table~\ref{tab:repair-results}) are evaluated: 

\begin{itemize}
    \item \emph{\textsc{KMeans} - BoW}: Uses \textsc{KMeans} and four different bag of words (BoW) based on \AAST, syntax, and invariants (lines 2--5 in Table~\ref{tab:repair-results});
    \item \emph{Closest Program (\textsc{KMeans}) - \AAST + Invs}: Uses the closest program (see Section~\ref{sec:implementation-closest-prog}) using the \AAST+ Invs BoW, from a set of clusters' representatives using \textsc{KMeans} (line 6);
\end{itemize}

Table~\ref{tab:repair-results} presents the overall repair evaluation on 319 incorrect submissions since  \Clara's repair algorithm does not support the C implementation of 101 incorrect submissions (24.05\% of the instances). Entries in bold correspond to the highest rate of submissions repaired, the lowest percentage of structural mismatch errors, or the lowest rate of timeouts, i.e., executions that did not repair a program using a timeout of 10 minutes (600s).

One can see in line 1 (Table~\ref{tab:repair-results})  that \Clara, using its own clusters, can only repair 229 (around 72\%) of the incorrect submissions and shows the largest percentage of instances that were not repaired due to timeout (20.69\%).
Secondly, the configuration using \InvAAST's \textsc{KMeans} and the BoW based on \AAST and Invariants achieved the highest score, repairing 84.95\% of the incorrect submissions. 
Furthermore, the BoW based only on invariants has the highest percentage of structural mismatch (11.91\%), which may be explained by \Clara's inability to use a program with a different control flow in the repair process. Using only invariants on a vector representation helps clustering programs with similar semantics, although it does not take into account the programs' structure (control flow). Hence, a higher rate of structural mismatch is observed.

Since the BoW based on \AASTs and program invariants achieved the best results both in the program clustering experiment (see Section~\ref{sec:results-clustering}) as well as when repairing submissions (lines 2-5 in Table~\ref{tab:repair-results}), we opted to use only this BoW when finding the closest correct program (line 6, Table~\ref{tab:repair-results}). 
Regarding the use of just one correct solution to fix an incorrect submission, the \emph{Closest Program (\textsc{KMeans})} approach did not achieve better results than using the set of clusters' representatives.

We have also analyzed the closest program technique using all submissions, i.e., use the closest program among all submissions (no clustering step). This approach, the \emph{Closest Program (All Submissions)}, was able to repair 86.5\% of the submissions. The number of timeouts in this approach and using \textsc{KMeans} was similar. Once again, this high rate of repaired programs (86.5\%) may be explained by \Clara's strict requirements for both programs, the program being repaired and the correct program used by the repair process, to have the same control flow. Therefore, when \InvAAST finds the closest program among all submissions instead of using clusters, \InvAAST has a more diverse collection of programs' structures. Consequentially, the \emph{Closest Program (All Submissions)} approach also achieved the lowest score of structural mismatch errors (only 9.4\%).
Although we would like to draw the reader's attention to the difference between the number of submissions repaired using \textsc{KMeans} (85\%) or using the closest correct program (86.5\%), which is less than 2\%. Furthermore, the computation to find the closest correct program among all correct submissions can only be done online since it requires the student's incorrect program. On the other hand, the computation of the \textsc{KMeans} clusters can be done offline since it only requires past students' correct submissions. In this evaluation, this is not a concern since each \IPA has at most a hundred correct submissions. However, in a large-scale MOOC with thousands of correct submissions per exercise, the process of finding the closest correct program among all of the submissions may become impractical to compute in a short period of time.

\begin{figure}[t!]
        \centering
        \scalebox{0.75}{\includegraphics[width=\textwidth]{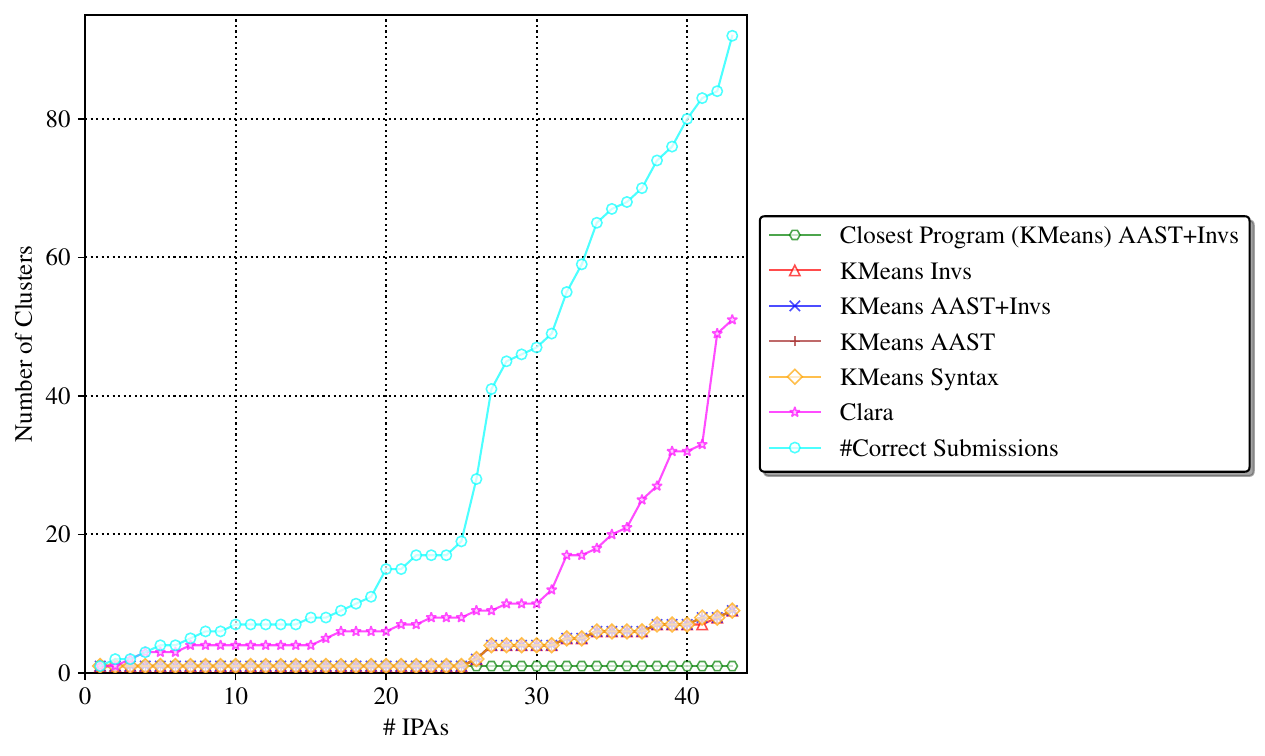}}
        \caption{Cactus plot - The number of clusters generated by each clustering technique for 43 different \IPAs.}
        \label{fig:num_clusters}
\end{figure}

\subsubsection{Number of Clusters}
\label{sec:results-repair-clustering-num-clusters}

Figure~\ref{fig:num_clusters} illustrates a cactus plot detailing the number of clusters generated by each clustering technique for each of the 43 different \IPAs used. One can see that \Clara generates an enormous quantity of clusters, almost half of the correct submissions of each \IPA. This large number of clusters is explained by \Clara's strict clustering method, which does not allow two programs to be in the same cluster if there is no exact match between both programs' control flows. 
Furthermore, even if two programs are semantically equivalent and share the same control flow but one of them uses a for-loop and the other uses a while-loop, then \Clara assigns these two programs to different clusters.

\InvAAST produces $K$ clusters, which in this experiment is always set to 10\% of the number of correct submissions of each exercise. The technique that uses the closest correct program has only a single cluster which is the closest correct program.
This evaluation of the number of clusters used by each approach allows us to observe that \Clara produces a large number of clusters, resulting in a detriment of performance. In contrast, our approach can generate fewer clusters resulting in a more effective repairing process.

\begin{example}
    The following two programs are correct implementations for the \IPA where students are asked to print the maximum value among three given numbers. These programs are clustered together using \InvAAST (\AASTs + Invariants) because their \AASTs are quite similar, and their sets of program invariants are identical.
    In contrast, \Clara assigns these two programs to different clusters because they have a different number of variables.
    Therefore, because \Clara is highly strict in comparing programs, it generates an excessive number of program clusters.
    
\begin{minipage}[t]{0.5\columnwidth}
\begin{code}
\label{code:prog_statements}
\begin{minted}[escapeinside=||,tabsize=1,obeytabs,xleftmargin=5pt,linenos]{C}
int main(){
    int n1, n2, n3, max;
    scanf("%d%d%d", &n1, &n2, &n3);
    max = n1;
    if(n2 > max){
        max = n2;
    }
    if(n3 > max){ 
        max = n3;
    }
	
    printf("%d\n", max);
    return 0;
}

\end{minted}
\end{code}
\end{minipage}
\begin{minipage}[t]{0.45\columnwidth}
\begin{code}
\label{code:statements_relaxed}
\begin{minted}[escapeinside=çç,tabsize=1,obeytabs,xleftmargin=5pt,linenos]{C}
int main(){
    int res,b,c;
    scanf("%d%d%d", &res, &b, &c);

    if (b > res){
        res = b;
    }
    if (c > res){
        res = c;
    }
   
    printf("%d\n", res);
    return 0;
}
\end{minted}
\end{code}
\end{minipage}

\end{example}

\subsubsection{CPU time}
\label{sec:results-repair-clustering-cpu-time}
\begin{figure}[t!]
    \centering
    \scalebox{0.75}{\includegraphics[width=\textwidth]{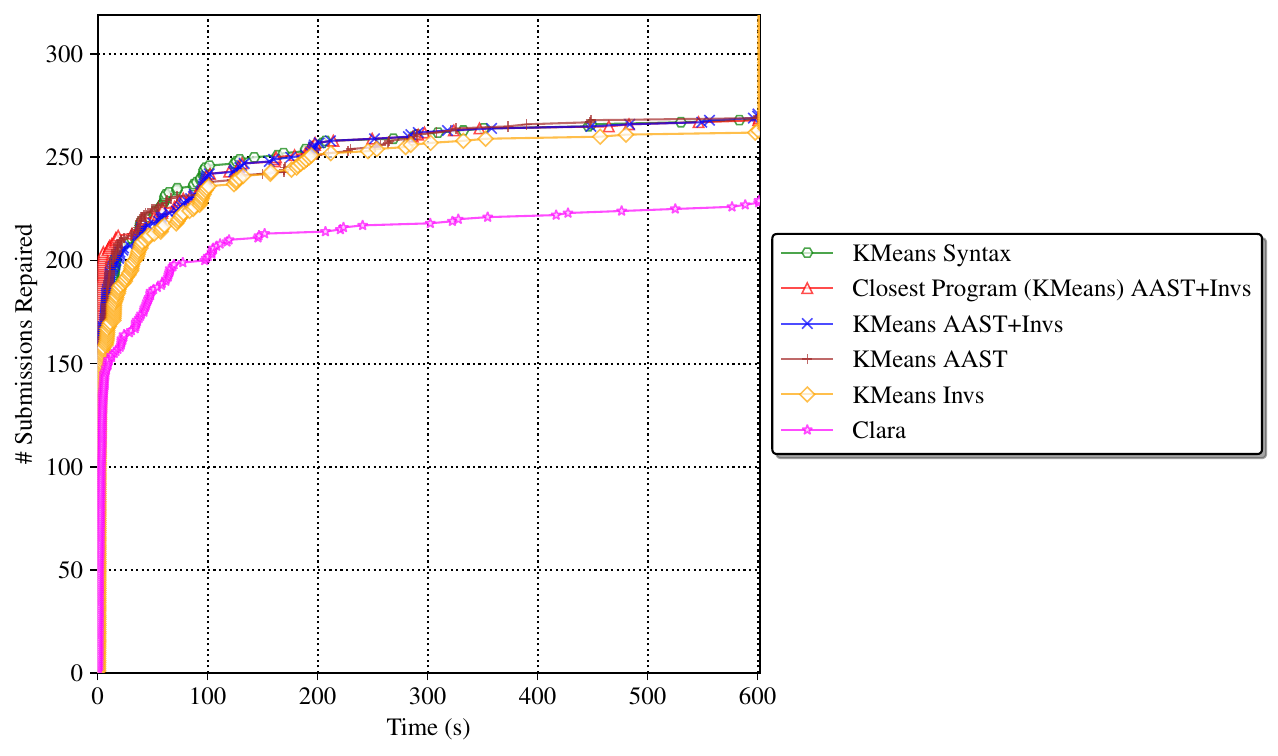}}
    \caption{Cactus plot - Time Performance (timeout=600s).}
    \label{fig:cpu_time}
\end{figure}

Regarding the time performance of each clustering technique, Figure~\ref{fig:cpu_time} shows a a cactus plot that presents the CPU time spent on repairing each program ($x$-axis) against the number of repaired programs ($y$-axis) using different clustering techniques.
The legend in this plot is not sorted.

\begin{figure}[t!]
\centering
\scalebox{0.5}{
\includegraphics[width=\textwidth]{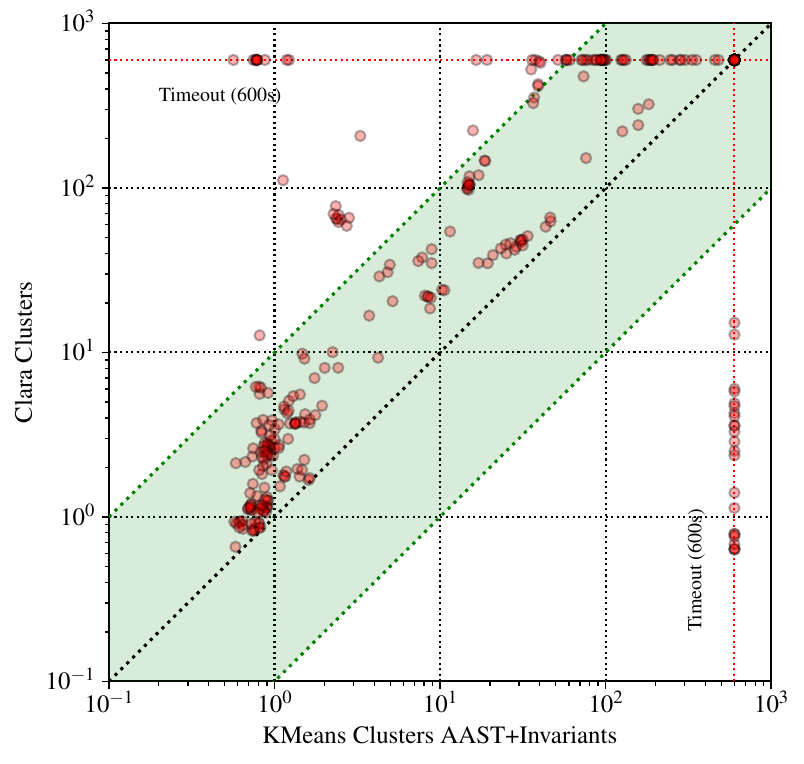}}
 \caption{Scatter plot - Time Performance (timeout=600s) - \Clara VS \InvAAST (\textsc{KMeans} w/ \AAST+ Invariants)}
 \label{fig:scatter-cpu-time}
\end{figure}

One can clearly see a gap between \Clara's time performance and \InvAAST's (considering any clustering approach). For example, after 100 seconds \Clara using its own clusters, can only repair around 200 programs while using our clusters, it can repair around 230/250 programs. Furthermore, Figure~\ref{fig:scatter-cpu-time} presents a scatter plot comparing the CPU time spent using \Clara's clusters against the \textsc{KMeans} \AAST + Invariants clusters.
Each point in this plot represents a program where the $x$-value (resp. $y$-value) is the CPU time spent to repair the program using the \textsc{KMeans} \AAST + Invariants Clusters (resp. \Clara's clusters). 
If a point is above the diagonal, then it means that our clusters outperformed \Clara's clusters since using our clusters \Clara is able to repair each program above the diagonal faster than using its own clusters. Thus, if we consider the programs repaired by both clustering methods, using our clusters is always faster than using \Clara's.

There are two main reasons for \Clara's time performance. Firstly, \Clara generates a significantly larger number of clusters compared to \InvAAST (see Figure~\ref{fig:num_clusters}). Consequently, \Clara needs to compute a set of repairs for each cluster's representative, resulting in more time spent in the repair process due to the larger number of clusters. Secondly, as explained in Section~\ref{sec:clara}, \Clara maintains a \texttt{json} file containing data (e.g., expressions) of all the programs belonging to the cluster. During its repair algorithm, \Clara takes advantage of all this data to repair a given submission, leading to more time spent on the repair process.

\begin{figure}[t!]
    \centering
     \scalebox{0.75}{\includegraphics[width=\textwidth]{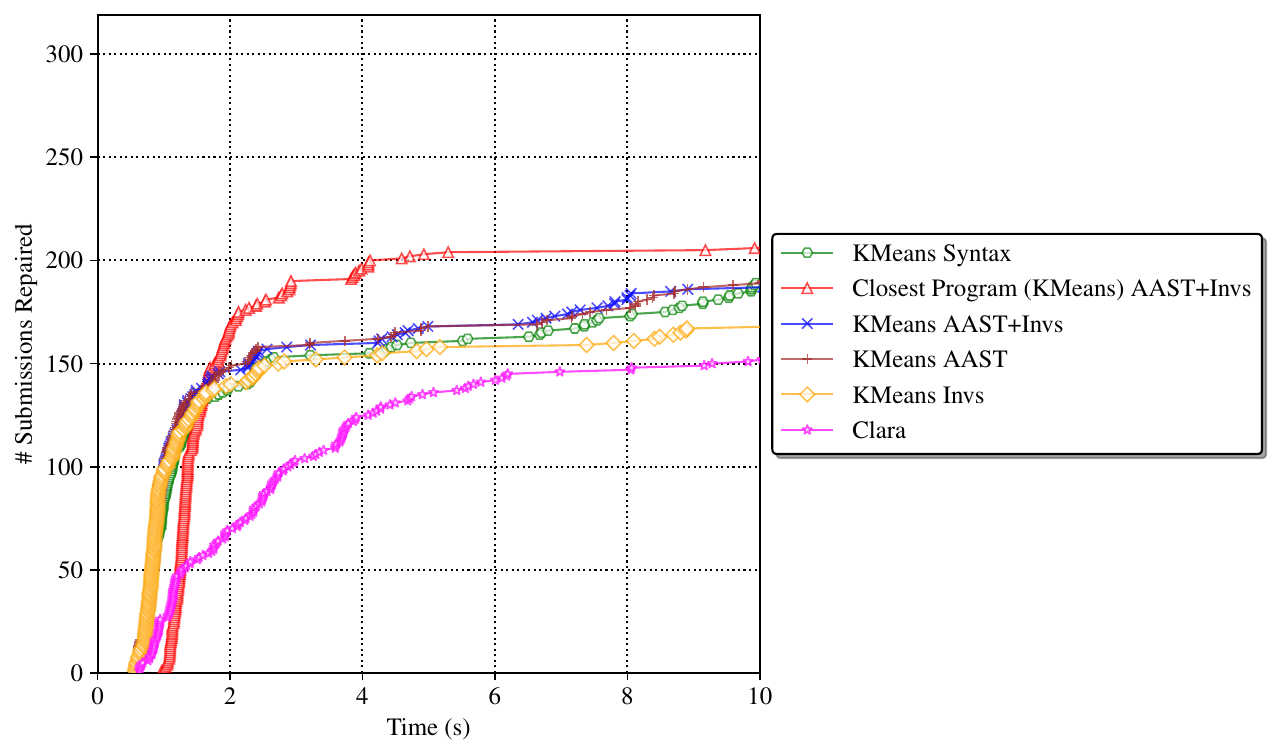}}
     \caption{Cactus plot - Time Performance (timeout=10s).}
     \label{fig:cpu_time_10s}
\end{figure}

The main goal of educational program repair frameworks is to provide real-time feedback to students on how they should repair their submissions. In this evaluation, we used a timeout of 10 minutes. However, a student expects a faster result. Therefore, Figure~\ref{fig:cpu_time_10s} shows another cactus plot that shows the time performance of the clustering approaches, although with a timeout of 10 seconds. One can see that after 10 seconds, \Clara using its own clusters, can only repair around 150 submissions~(47\%). On the other hand, using \InvAAST, \Clara can repair around 200 submissions~(63\%). Furthermore, one can also verify that after 2 seconds \Clara can only repair around 75 programs using its own cluster while using our clusters \Clara is able to repair around 150 programs.

\subsubsection{Program invariants of incorrect submissions}
\label{sec:results-repair-composed}
We have also tried the method \emph{Closest Program (\textsc{KMeans})} with \InvAAST not taking into account incorrect submissions' invariants to check if incorrect submissions' sets of invariants had a negative impact on program representations. First, \InvAAST clustered all the correct programs considering their \AASTs and their sets of invariants. Secondly, to find the closest correct program to an incorrect submission \InvAAST used only the \AAST BoW.
However, this combined approach of clustering with one BoW (\AASTs + invariants) and calculating the programs' distances with another (\AASTs only) was only able to repair 269 submissions (84\%). Thus, according to this experiment, incorrect submissions' sets of program invariants do not cause negative effects on program representations.

\mbox{}\hfill

To summarize, in Section~\ref{sec:results-clustering}, we used \InvAAST to cluster different \IPAs correct students' submissions. The obtained results support that this work's novel program representation based on a program's \AAST and invariants performs better when compared to representations solely based on a program's code, \AST, or set of program invariants. 

Furthermore, in Section~\ref{sec:results-repair}, we integrated \InvAAST into \Clara to evaluate our tool's performance when integrated into a clustering-based framework for repairing \IPAs. This study shows that \InvAAST significantly increases the performance of the state-of-the-art clustering technique by allowing \Clara to repair more student submissions and doing so notably faster.

\subsection{Threats to Validity}
\label{sec:threats2validity}

This work relies on \Daikon to compute dynamically-generated likely invariants. Using another tool to detect likely invariants may produce different results.
Dy\-na\-mi\-ca\-lly-ge\-ne\-ra\-ted likely invariants depend highly on the test suite used for each programming exercise. Therefore, using a different set of input-output tests may produce different sets of program invariants for each student submission.

Furthermore, our evaluation of \InvAAST exclusively employed small imperative programs commonly encountered in \IPAs. The application of \InvAAST to more intricate programs is deferred to future investigations, although we anticipate no major scalability issues with larger datasets or more complex programs when utilizing \InvAAST. Additionally, it is worth noting that the number of clusters expands based on the diversity of semantic and syntactic implementations for each \IPA. In theory, as the complexity of \IPAs increases, we can expect a greater diversity of implementations. Nevertheless, the scalability of \InvAAST should remain unaffected, and there is no imposed limit on the maximum number of supported \IPAs.

While our evaluation focused solely on C programs, our clustering methodologies are agnostic and can be adapted to other programming languages. This adaptation entails substituting (1) the variable renaming module with one compatible with the target language and (2) \InvAAST's invariant detection module (\Daikon) with another suitable program invariant detector tailored for the desired language.

A different repair tool may produce different results since the repair process may differ. For example, another tool may support C features that \Clara currently does not support or the opposite.
Only small imperative programs, usually found in \IPAs, were used to evaluate \InvAAST. Using \InvAAST on more complex programs is left for future work, although we anticipate no scalability issues with larger programs.

\InvAAST could provide an intriguing perspective on utilizing \AASTs and program invariants to identify potential similarities/copies among students' submissions. However, implementing this would require some time to adapt the plagiarism detection tool. Additionally, it is worth noting that \textsc{Moss} is not open-source.

In this paper, we compared \InvAAST against \Clara's clustering method since, to the best of our knowledge, \Clara is currently the only publicly available state-of-the-art clustering-based repair tool for repairing \IPAs. However, evaluating \InvAAST on other repair frameworks would be valuable. Unfortunately, we could not find public implementations for the other tools~\cite{FAPR-corr-21, pacon-oopsla-21, semCluster-pldi19, sarfgen} nor the datasets of \IPAs used for their evaluation, except for the ITSP dataset~\cite{asr-for-ITSP}.

\section{Related Work}
\label{sec:related-work}

\emph{Program clustering} has also been used to find different semantic solutions for a given programming exercise~\cite{overcode15, pacon-oopsla-21, equivalence-funct-prog-oopsla20}. \textsc{PaCon}~\cite{pacon-oopsla-21} clusters programming assignments based on their symbolic analysis. \textsc{PaCon} clusters two submissions together if their path conditions are equivalent. \textsc{PaCon} only takes into consideration a program's semantics, while \InvAAST also considers a program's structure. \textsc{Overcode}~\cite{overcode15} lets the user visualize and explore different implementations for the same exercise.
\textsc{CodeBERT}~\cite{codeBERT-emnlp22}, \textsc{code2seq}~\cite{code2seq-iclr19} and other deep learning models~\cite{iclr21-source-code-representation} build vector representations of programs by training machine learning models using the programs' code and ASTs. However, unlike \InvAAST, these techniques only consider the programs' syntax, not their semantics.

\textsc{CoderAssist}~\cite{coderAssist-fse16} is a counter-example guided feedback generation tool that provides feedback on student implementations of \emph{dynamic programming algorithms}.
\textsc{CoderAssist} starts by clustering both correct and incorrect programs based on these dynamic programs' syntactic features.
Afterward, \textsc{CoderAssist} generates feedback for a buggy program by calling an SMT solver, using a counterexample obtained from an equivalence check against a correct implementation in the same cluster.
However, \textsc{CoderAssist} only works for dynamic programs since the implemented feature extraction procedure searches for syntactic features of dynamic programming algorithms. Thus, this tool cannot be used for clustering our dataset of \IPAs. \citet{rocha2023helping} introduce a framework to improve how teaching assistants provide feedback in introductory programming courses, helping them understand different feedback approaches and resources. This framework offers structured guidance for pedagogical decision-making regarding adaptive feedback.

\textsc{CodeHound}~\cite{codehound-splash22} is a system that automatically tracks pedagogical code dependencies by using static analysis to detect function introductions and reuse throughout an entire course. It offers instructors assistance in creating new content, collaborating on content refactoring, and estimating future course change costs. Furthermore, \textsc{ODS}~\cite{tse22-ODS} is a system for detecting overfitting patches in automatic program repair. ODS utilizes supervised learning with \AST level code features and patch correctness labels to automatically learn a probabilistic model, which can then classify new program repair patches. Moreover, \textsc{CLACER}~\cite{compsac21-CLACER} is a neural network model designed to classify compilation errors by proposing categories based on program tokens, aiming to improve localization effectiveness and prediction performance, thereby enhancing the students' learning process.

\emph{Code search techniques}~\cite{code-search-icse09-reiss, sosRepair-tse19,searchRepair,fse-nier12-semantic-search, tsem16-code-search-io-queries, oopsla19-aroma-code-recomendation, fse21-cross-language-code-search} are another family of semantic program repair techniques. Code search uses a specification (e.g., input-output tests) to find code in a large repository of code snippets that satisfies that specification. To repair an incorrect program, semantic code search methods do not find the closest correct implementation for the same \IPA or use a reference implementation provided by the lecturer. Instead, these methods search for code fragments of several correct programs to repair a given incorrect submission.
These methods have no knowledge about the program's structure and where that code fragment came from. Therefore, there is no guarantee that the set of repairs proposed by these tools is the minimal set required to fix a program.

\emph{Solution-driven program repair} tools use one reference implementation to repair a given incorrect submission~\cite{verifix, refactory, autograder}. \textsc{AutoGrader}~\cite{autograder} finds potential path differences between the executions of a student's submission and a reference implementation using symbolic execution. Then, \textsc{AutoGrader} provides feedback to students using counter-examples for each path difference found.
\textsc{Verifix}~\cite{verifix} aligns an incorrect program with the reference solution into an automaton. Then, using that alignment relation and MaxSMT solving, \textsc{Verifix} proposes fixes to the incorrect program.
\textsc{TEGCER}~\cite{tegcer-ase19} is an automated feedback tool
for novice programmers. This tool uses supervised learning to match \emph{compilation errors} in new code submissions with pre-existing errors submitted by previously enrolled students.
\textsc{TEGCER} only works for syntactic errors. However, similarly to \InvAAST, \textsc{TEGCER} also performs variable renaming to each program. This tool renames the variables with their generic types using the \textsc{LLVM}~\cite{llvm-cgo04}, a standard static analysis tool.

Regarding \emph{clustering-based program repair tools}~\cite{semCluster-pldi19, FAPR-corr-21, sarfgen, clara}, \Clara~\cite{clara} was already described, and differences were highlighted in previous sections. More recently, \citet{its22-improving-clara-matching-procedure, jss-ChowdhuryCR24-ClaraUpdate} improved \Clara's matching algorithm to a new graph matching algorithm that is more relaxed in terms of control flow restrictions. However, this new graph-matching algorithm only works for Python programs.
\textsc{SarfGen}~\cite{sarfgen}, used for C\# programs, creates program embeddings based on the programs' ASTs~\cite{deckard}. Then, given an incorrect program, finds the closest correct submission using those embeddings and tries to repair the program by aligning the variables in both programs. Thus, unlike \InvAAST, \textsc{SarfGen} only considers a program's structure (AST) and not its semantics during the clustering process.

\textsc{SemCluster}~\cite{semCluster-pldi19} clusters programs based on their control and data flow features. \textsc{SemCluster} creates vector representations, \emph{program features vectors} (PFV), for each program using a test suite. This PFV takes into account control flow features as well as data flow features. For each program, \textsc{SemCluster} counts the number of times each control flow path is used in each test and builds a vector with this data. Afterward, \textsc{SemCluster} builds another vector containing the data flow features, i.e., the number of occurrences of consecutive values a variable takes during its lifetime. Finally, \textsc{SemCluster} merges these two feature vectors into a single vector, the PFV. Similar to \InvAAST, \textsc{SemCluster} uses the \textsc{KMeans} clustering algorithm. However, unlike \InvAAST, which uses each program's \AAST, \textsc{SemCluster} does not account for any syntactic features. Furthermore, \textsc{SemCluster} tries to capture the semantics of a program by counting the number of different values each variable takes. On the other hand, \InvAAST considers the program's set of invariants which can be more robust and independent of the test suite used.
Some research has been conducted regarding \emph{the use of invariants to promote patch diversity} or to help with patch selection on a search-based program
repair~\cite{ase19-weimer-understanding-patches-through-invariants, gi-icse19-invariants-diversity-search-based-repair, ibf20-yang-invariants-difference-patches}.

\section{Conclusions and Future Work}
\label{sec:conclusion}

In the context of introductory programming assignments (\IPAs) in university courses or Massive Open Online Courses (MOOCs), it is possible to collect a large number of correct implementations for the proposed \IPAs. Hence, when a student submits an incorrect program, one can take advantage of previously correct submissions to automatically suggest repairs that help the student. However, it is not feasible to analyze all possible previous correct submissions to find an appropriate reference implementation for the repair tool. Therefore, clustering is often used to identify similar program implementations. Afterward, the automated repair tool analyses a single reference program from each cluster to find the most suitable correction to the student's incorrect submission.

This work proposes \InvAAST, a novel approach for program clustering based on their semantic and syntactic features. \InvAAST uses \AASTs and invariant-based program representations to
distinguish small imperative programs according to their semantics (invariants) and structure (\AAST).
Results show that the proposed \AASTs and invariant-based representation improve upon syntax-based representations when performing program clustering on several correct student submissions for different programming exercises.
Additionally, given an incorrect student submission and a set of correct students' submissions, \InvAAST can also find the closest correct submission to the faulty program using \InvAAST's program vector representations.

Furthermore, \InvAAST has also been integrated into a state-of-the-art clustering-based program repair framework to evaluate the proposed clustering techniques for repairing \IPAs. This evaluation showed that \InvAAST outperforms the current state-of-the-art clustering method used in clustering-based program repair. Using \InvAAST, the automated repair tool \Clara can repair significantly more \IPAs, around 13\%, with a better time performance and with a smaller number of program clusters.

To conclude, \InvAAST is a program clustering tool based on programs' invariants and \AASTs. \InvAAST can be used: (1) to cluster semantically equivalent implementations for programming exercises; 
(2) by any clustering-based program repair tool; and (3) by any program repair framework that requires a single reference implementation (\InvAAST's closest correct program).

As future directions, we propose to evaluate the new program representations described in this paper (i.e., using \AASTs and invariants) as program encodings to use on deep learning models on several tasks such as fault localization~\cite{jss23-vsusfl} or program synthesis~\cite{master-thesis-pedro}. As we expand to consider more complex programs, we plan to evaluate \InvAAST on clustering-based repair tools focused on repairing industrial software.
Finally, \InvAAST will be evaluated on other clustering-based program repair frameworks with a more permissive repair algorithm than \Clara. Moreover, it could also be interesting to explore using a neural architecture to learn deep representations of the features built up in our model, as the current Bag of Words (BoW) approach may lead to a significant loss of information.

\section*{Acknowledgments}
This research was supported by Fundação para a Ciência e Tecnologia (FCT) through grant SFRH/\-BD/\-07724/\-2020 (DOI: 10.54499/\-2020.07724.BD) and projects UIDB/50021/2020 (DOI: 10.54499/\-UIDB/\-50021/\-2020), PTDC/\-CCI-COM/\-2156/2021 (DOI: 10.54499/\-PTDC/\-CCI-COM/\-2156/\-2021) and 2023.14280.PEX (DOI: 10.54499/2023.14280.PEX) and grant SFRH/\-BD/\-07724/\-2020 (DOI: 10.54499/\-2020.07724.BD).
This work was also supported by the MEYS within the program ERC CZ under the project POSTMAN no.~LL1902 and co-funded by the EU under the project \emph{ROBOPROX} (reg.~no. CZ\-.02.01.01/00/\-22\_008/0004590).



\bibliographystyle{cas-model2-names}

\bibliography{main}

\appendix

\section{Description of \benchmark}
\label{appendix:description-IPAs}
The set of \IPAs corresponds to three different lab classes of the introductory programming course to the C programming language at Instituto Superior Técnico, Universidade de Lisboa. Each lab class focuses on a different topic of the C programming language. 
Lab02 is described in Section~\ref{sec:lab02} deals with integers and input-output operations. Section~\ref{sec:lab03} presents Lab03, which focuses on loops and chars. Lastly, Section~\ref{sec:lab04} describes Lab04, where the students learn to use vectors and strings.
The textual description of each programming assignment can be found in the next sections. The input/output tests used to evaluate semantically the set of students' submissions can be found at the public GitHub~\href{https://github.com/pmorvalho/C-Pack-IPAs}{https://github.com/pmorvalho/C-Pack-IPAs} of \benchmark~\cite{C-Pack-IPAs_apr24}. Moreover, there is also a reference implementation for each \IPA in the public git repository that can be used by program repair frameworks that only accept a single reference implementation to repair incorrect programs.

\subsection{Lab02 - Integers and IO operations.}
\label{sec:lab02}
In Lab02, the students learn how to program with integers, floats, IO operations (mainly \texttt{printf} and \texttt{scanf}), conditionals (if-statements), and simple loops (for and while-loops).

\paragraph{\IPA\#1: Lab02 - Ex01.} Write a program that determines and prints the largest of three integers given by the user.

\paragraph{\IPA\#2: Lab02 - Ex02.} Write a program that reads two integers `N, M` and prints the smallest of them in the first row and the largest in the second.

\paragraph{\IPA\#3: Lab02 - Ex03.} Write a program that reads two positive integers `N, M` and prints "yes" if `M` is a divisor of `N`, otherwise prints "no".

\paragraph{\IPA\#4: Lab02 - Ex04.} Write a program that reads three integers and prints them in order on the same line. The smallest number must appear first.

\paragraph{\IPA\#5: Lab02 - Ex05.} Write a program that reads a positive integer `N` and prints the numbers `1..N`, one per line.

\paragraph{\IPA\#6: Lab02 - Ex06.}Write a program that determines the largest and smallest number of `N` real numbers given by the user. Consider that `N` is a value requested from the user.
The result must be printed with the command `printf("min: \%f, max: \%f\~n", min, max)`.
\emph{Hint: initialize the largest and smallest to the first read value.}

\paragraph{\IPA\#7: Lab02 - Ex07.} Write a program that asks the user for a positive integer `N` and prints the number of divisors of `N`. Remember that prime numbers have 2 divisors.

\paragraph{\IPA\#8: Lab02 - Ex08.}  Write a program that calculates and prints the average of `N` real numbers given by the user. The program should first ask the user for an integer `N`, representing the number of numbers to be entered. The real numbers must be represented
by float type. The result must be printed with the command `printf("\%.2f", avg);`.

\paragraph{\IPA\#9: Lab02 - Ex09.} Write a program that asks the user for a value `N` corresponding to a certain period of time in seconds. The program should output this period of time in the format `HH:MM:SS`.

\emph{Hint: use the operator that calculates the remainder of division (`\%`).}

\paragraph{\IPA\#10: Lab02 - Ex10.} Write a program that asks the user for a positive value `N`. The output should present the number of digits that make up `N` (on the first line), as well as the sum of the digits of `N` (on the second line). For example, the number 12345 has 5 digits, and the sum of these digits is 15.

\subsection{Lab03 - Loops and Chars.}
\label{sec:lab03}
In this lab, the students learn how to program with loops, nested loops, auxiliary functions, and chars.

\paragraph{\IPA\#11: Lab03 - Ex01.} Write a program that draws a square of numbers like the following using the function `void square(int N);`. The value of `N`, given by the user, must be greater than or equal to 2. The tab (character `'\t'`) must be used as the separator. The square shown is the example for `N = 5`.

\begin{center}
\emph{1 2 3 4 5}

\emph{2 3 4 5 6}

\emph{3 4 5 6 7}

\emph{4 5 6 7 8}

\emph{5 6 7 8 9}
\end{center}

\paragraph{\IPA\#12: Lab03 - Ex02.} Write a program that draws a pyramid of numbers using the `void pyramid(int N);` function. The value of `N`, given by the user, must be greater than or equal to 2. The space (character `' '`) must be used as the separator. The pyramid shown is the example for `N = 5`.

\begin{center}
\emph{        1        }
        
\emph{      1 2 1      }
      
\emph{    1 2 3 2 1    }
    
\emph{  1 2 3 4 3 2 1  }
  
\emph{1 2 3 4 5 4 3 2 1}
\end{center}

\paragraph{\IPA\#13: Lab03 - Ex03.} Write a program that draws a cross on diagonals using the `void cross(int N);` function. The asterisk (`'*'` character) must be used to draw the cross; hyphen (`'-'` character) must be used as the separator. The crosses shown are the examples for `N = 3` and `N=8`.

\begin{center}
* - *

- * -

* - *
\end{center}

\begin{center}
* - - - - - - *

- * - - - - * -

- - * - - * - -

- - - * * - - -

- - - * * - - -

- - * - - * - -

- * - - - - * -

* - - - - - - *
\end{center}

\paragraph{\IPA\#14: Lab03 - Ex04.} Write a program that reads a sequence of numbers separated by spaces and newlines,
and print the same string, but the numbers in the output should not contain 0 at the beginning, eg `007` should print `7`.
The exception is the number 0, which should be printed as 0.
The string in the input ends with `EOF`.

\emph{Warning: Number values may be greater than the maximum value of type `int` or any primitive type in C.}

\emph{Hint: the `int getchar()` function can be used to read a character.}

\paragraph{\IPA\#15: Lab03 - Ex05.} Write a program that reads a sequence of messages and prints them out, one per line.
Each message is delimited by quotation marks (character `"`). The message can contain an "escape sequence" - the character loses special meaning if it is preceded by the character `\` (backslash). For example, the input `"a\" foo\~bar\""` matches the message `a"foo\~bar"`. So the backslash allows you to include quotes in the message just like the backslash itself.

\emph{Hint: use the concept of state as we did in the word counter in the theoretical class.}

\paragraph{\IPA\#16: Lab03 - Ex06.} Write a program that reads a positive integer from the input (such as a sequence of characters up to 100 chars) and that decides whether the number read is divisible by 9.
If the number is divisible by 9, the program should print the message `yes`, and should print `no` otherwise.

\emph{Warning: Number values can be greater than the maximum value of type `int` or any primitive type in C.}

\emph{Hint: A number is divisible by 9 if and only if the sum of its digits is divisible by 9.}

For example, the sum of the digits of the number 729 is 18, so it is divisible by 9.
The fact can be seen from the following equation: 7 x 100 + 2 x 10 + 9 = (7 x 99 + 7) + (2 x 9 + 2) + 9.

\paragraph{\IPA\#17: Lab03 - Ex07.} Write a program that takes a sequence of numbers and operators (`+`, `-`) representing an arithmetic expression and returns the result of that arithmetic expression.
The string in the input ends with `\~n`.
You can assume that every two numbers are always separated by `space, operator, space`, i.e., `'op'`, for either of the 2 operators above.
Example: Input `70 + 22 - 3` should return `89`.

\emph{Hint: You should start by converting a sequence of digits (characters) to an integer.}

\subsection{Lab04 - Vectors and Strings.}
\label{sec:lab04}
In this lab, the students learn how to program with integers arrays, and strings.

\paragraph{\IPA\#18: Lab04 - Ex01.} Write a program that asks the user for a positive integer `n < VECMAX`, where `VECMAX=100`. Then read `n` positive integers. In the end, the program should write a graphical representation of the values read as follows. The graph shown is the example for `n = 3` and values `1 3 4`.

\begin{center}
\emph{*}

\emph{***}

\emph{****}
\end{center}

\paragraph{\IPA\#19: Lab04 - Ex02.} Write a program that asks the user for a positive integer `n < VECMAX`, where `VECMAX=100`. Then read `n` positive integers. In the end, the program should write a graphical representation of the values read as follows. The graph shown is the example for `n = 3` and values `1 3 4`.

\begin{center}
\emph{***}

\emph{**}

\emph{**}

\emph{*}
\end{center}

\paragraph{\IPA\#20: Lab04 - Ex03.} Write a program that asks the user for a positive integer `n < VECMAX`, where `VECMAX=100`. Then read `n` positive integers. In the end, the program should write a graphical representation of the values read as follows. The graph shown is the example for `n = 3` and values `1 3 4`.

\begin{center}
\emph{*}

\emph{**}

\emph{**}

\emph{***}
\end{center}

Consider that in the following \IPAs, all strings have a maximum of `MAX = 80` characters (including the end-of-string character).

\paragraph{\IPA\#21: Lab04 - Ex04.} Write a program that reads a word from the terminal and checks whether the word is a palindrome or not. A word is a palindrome if it is spelled the same way from left to right and vice versa (eg "AMA" is a palindrome). If the word is a palindrome, the program should print the value `yes`, and `no` if not.

\emph{Hint: You can use `scanf("\%s", s)` to read a word. Note that the string `s` does not ask for `\&` in `scanf`.}

\paragraph{\IPA\#22: Lab04 - Ex05.} Write a program that reads characters from the keyboard, character by character until it finds the character `\~n` or EOF and writes the line read to the terminal. Implement the `int leLinha(char s[])` function, which reads the line into the string `s` and returns the number of characters read.
\emph{Hint: After solving this exercise, try using the `fgets` command.}

\paragraph{\IPA\#23: Lab04 - Ex06.} Write a program that reads a line from the terminal (use the function from the previous exercise) and writes the same text to the terminal but with the lowercase letters replaced by the respective uppercase letters. Implement the `void uppercase(char s[])` function. \emph{Note: Remember that the string `s` is changed by the `uppercase` function.}

\paragraph{\IPA\#24: Lab04 - Ex07.} Write a program that reads a line and a character and writes to the terminal the same line where all occurrences of the character were removed. Implement the `void eraseCharacter(char s[], char c)` function that erases the character `c` from the string `s`.

\paragraph{\IPA\#25: Lab04 - Ex08.} Write a program that reads two integers in decimal representation and prints the larger of those two numbers. You can assume that the two numbers have the same number of digits and a maximum of 100 characters.

\emph{Note: The numbers may be too large to be stored in a `long long` variable, for example `998888\-888888\-8888\-88887` and `99888\-888888\-888888\-88888`.}

\section{Use Case \#1: Clustering \IPAs}
\label{appendix:clustering}

\subsection{Clustering Accuracy}
\label{appendix:clustering-acc}

\begin{figure}[t!]
    \centering
    \scalebox{1}{
    \includegraphics{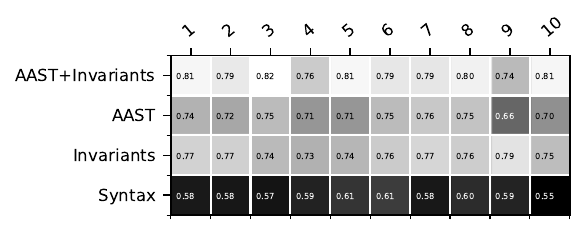}
    }
    \caption{The values for cluster accuracy using the \textsc{MiniBatch KMeans} algorithm on each program representation after ten different runs, each run using a different seed.}
    \label{fig:acc-mbk}
\end{figure}

Figure~\ref{fig:acc-mbk} shows a matrix with the different values of the cluster accuracy using the \textsc{MiniBatch KMeans} algorithm on each program representation using ten different seeds. Each entry is highlighted accordingly to its value. The lowest value is highlighted in black, and the highest is highlighted in white. Intermediate values are highlighted in different shades of grey, depending on how far they are from the lowest value.

\begin{figure}[t!]
    \centering
    \scalebox{1}{
    \includegraphics{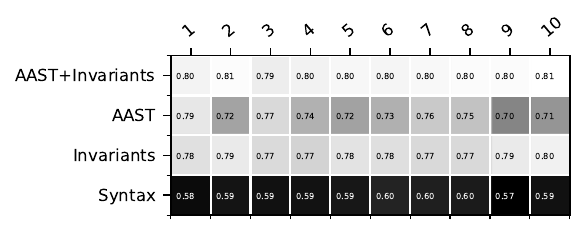}
    }
    \caption{The values for cluster accuracy using the \textsc{Birch} algorithm on each program representation after ten different runs, each run using a different seed.}
    \label{fig:acc-b}
\end{figure}

Secondly, Figure~\ref{fig:acc-b} shows a matrix with the different values of the cluster accuracy using the \textsc{Birch} algorithm on each program representation using ten different seeds.

\begin{figure}[t!]
    \centering
    \scalebox{1}{
    \includegraphics{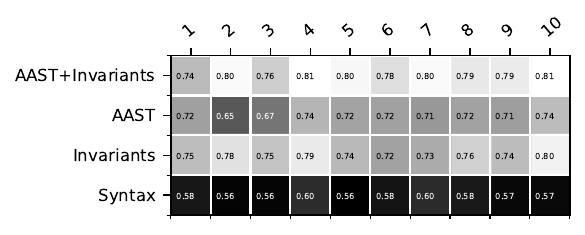}
    }
    \caption{The values for cluster accuracy using the \textsc{Gaussian Mixture} algorithm on each program representation after ten different runs, each run using a different seed.}
    \label{fig:acc-gm}
\end{figure}

Lastly, Figure~\ref{fig:acc-gm} shows a matrix with the different values of the cluster accuracy using the \textsc{Gaussian Mixture} algorithm on each program representation for ten different seeds.

\subsection{Other Evaluation Metrics}
\label{appendix:clustering-metrics}

In this section, we present other clustering evaluation metrics for the \textsc{KMeans} algorithm, such as: the \emph{Rand index}, the \emph{adjusted Rand index}, the \emph{normalized mutual information}, the \emph{adjusted mutual information}, the \emph{Fowlkes–Mallows index}, the \emph{completeness score}, the \emph{homogeneity score}, and the \emph{V measure}.

\paragraph{\textbf{Rand Index.}} The \emph{Rand index} measures the similarity of the two assignments, ignoring permutations~\cite{rand1971objective}. The Rand index is given by the following equation~\ref{eq:rand-index}:

\begin{equation}
RI={\frac {TP+TN}{TP+FP+FN+TN}}
\label{eq:rand-index}
\end{equation}

\begin{figure}[t!]
    \centering
    \scalebox{1}{
    \includegraphics{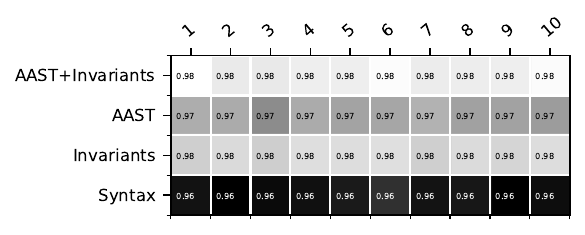}
    }
    \caption{The values for the Rand index using the \textsc{KMeans} algorithm on each program representation after ten different runs, each run using a different seed.}
    \label{fig:rand_index}
\end{figure}

Figure~\ref{fig:rand_index} presents the values for the Rand index using the \textsc{KMeans} algorithm on each program representation after ten different runs.

\paragraph{\textbf{Adjusted Rand Index.}} The \emph{adjusted Rand index} is the corrected-for-chance version of the Rand index since the Rand index does not guarantee that random label assignments will get a value close to zero~\cite{adj_rand_index}. The adjusted Rand index is given by equation~\ref{eq:adj_rand-index}, where $TP$ is the number of true positives, $TN$ is the number of true negatives, $FP$ is the number of false positives, and $FN$ is the number of false negatives.

\begin{equation}
ARI={\frac {RI - E[RI]}{max(RI) - E[RI]}}
\label{eq:adj_rand-index}
\end{equation}

\begin{figure}[t!]
    \centering
    \scalebox{1}{
    \includegraphics{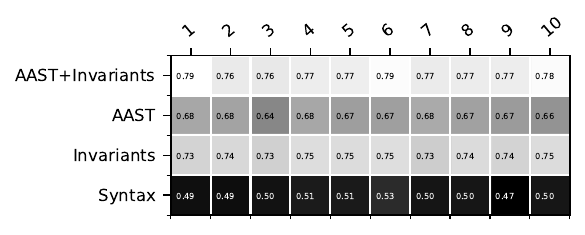}
    }
    \caption{The values for the adjusted Rand index using the \textsc{KMeans} algorithm on each program representation after ten different runs, each run using a different seed.}
    \label{fig:adj_rand_index}
\end{figure}

Figure~\ref{fig:adj_rand_index} presents the values for the adjusted Rand index using the \textsc{KMeans} algorithm on each program representation after ten different runs.

\begin{figure}[t!]
    \centering
    \scalebox{1}{
    \includegraphics{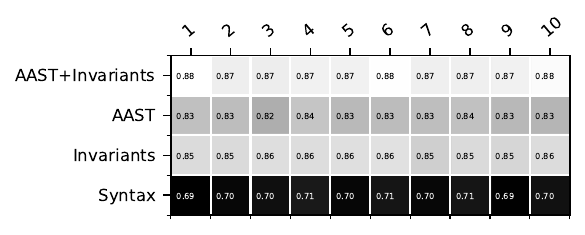}
    }
    \caption{The values for the normalized mutual information using the \textsc{KMeans} algorithm on each program representation after ten different runs, each run using a different seed.}
    \label{fig:nmi}
\end{figure}

\paragraph{\textbf{Normalized Mutual Information.}} The \emph{normalized mutual information} of two random variables is a measure of the mutual dependence between the two variables~\cite{mutual_information}.
Figure~\ref{fig:nmi} shows the values for the normalized mutual information using the \textsc{KMeans} algorithm on each program representation after 10 different runs.

\begin{figure}[t!]
    \centering
    \scalebox{1}{
    \includegraphics{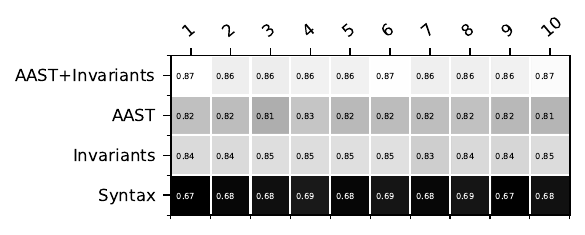}
    }
    \caption{The values for the adjusted mutual information using the \textsc{KMeans} algorithm on each program representation after ten different runs, each run using a different seed.}
    \label{fig:adj_mi}
\end{figure}

\paragraph{\textbf{Adjusted Mutual Information.}} The \emph{adjusted mutual information} corrects the effect of the agreement solely due to chance between clusterings, similar to the way the adjusted rand index corrects the Rand index~\cite{adj_mutual_information}. Figure~\ref{fig:adj_mi} presents the values for the adjusted mutual information using the \textsc{KMeans} algorithm on each program representation after ten different runs.

\paragraph{\textbf{Fowlkes–Mallows index.}} The \emph{Fowlkes-Mallows index} measures the similarity between two clusters. A high value for the Fowlkes–Mallows index indicates a great similarity between the clusters and the benchmark classifications~\cite{fowlkes1983method}.
The Fowlkes-Mallows index can be computed using equation~\ref{eq:fmi}, where $TP$ is the number of true positives, $FP$ is the number of false positives, and $FN$ is the number of false negatives. $TPR$ is the true positive rate, also called sensitivity or recall, and $PPV$ is the positive predictive rate, also known as precision.

\begin{equation}
    \label{eq:fmi}
    {\displaystyle FM={\sqrt {PPV\cdot TPR}}={\sqrt {{\frac {TP}{TP+FP}}\cdot {\frac {TP}{TP+FN}}}}}
\end{equation}

\begin{figure}[t!]
    \centering
    \scalebox{1}{
    \includegraphics{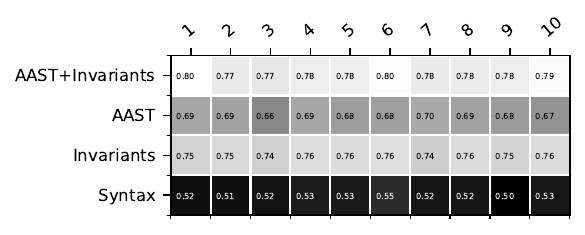}
    }
    \caption{The values for the Fowlkes–Mallows index using the \textsc{KMeans} algorithm on each program representation after ten different runs, each run using a different seed.}
    \label{fig:fmi}
\end{figure}

Figure~\ref{fig:fmi} shows the values for the Fowlkes–Mallows index using the \textsc{KMeans} algorithm on each program representation after ten different runs.

\begin{figure}[t!]
    \centering
    \scalebox{1}{
    \includegraphics{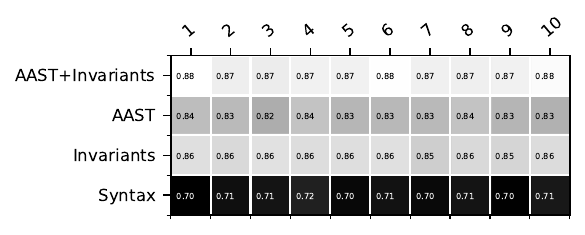}
    }
    \caption{The values for the completeness score using the \textsc{KMeans} algorithm on each program representation after ten different runs, each run using a different seed.}
    \label{fig:comp_score}
\end{figure}

\paragraph{\textbf{Completeness score.}} The \emph{completeness score} measure if all members of a given class are assigned to the same cluster~\cite{v-measure}. Figure~\ref{fig:comp_score} shows the values for the completeness score using the \textsc{KMeans} algorithm on each program representation after ten different runs.

\begin{figure}[t!]
    \centering
    \scalebox{1}{
    \includegraphics{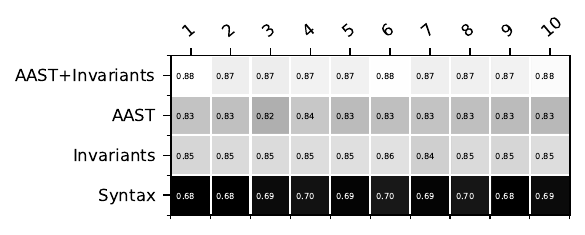}
    }
    \caption{The values for the homogeneity score using the \textsc{KMeans} algorithm on each program representation after ten different runs, each run using a different seed.}
    \label{fig:homo_score}
\end{figure}

\paragraph{\textbf{Homogeneity score.}} The \emph{homogeneity score} checks if each cluster contains only members of a single class~\cite{v-measure}.
Figure~\ref{fig:homo_score} presents the values for the homogeneity score using the \textsc{KMeans} algorithm on each program representation after ten different runs.

\begin{figure}[t!]
    \centering
    \scalebox{1}{
    \includegraphics{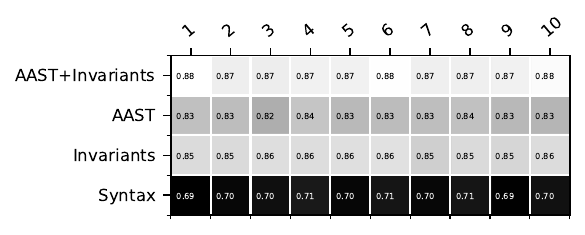}
    }
    \caption{The values for the V measure using the \textsc{KMeans} algorithm on each program representation after ten different runs, each run using a different seed.}
    \label{fig:v_measure}
\end{figure}

\paragraph{\textbf{V measure.}} The \emph{V measure} is the harmonic mean between the homogeneity and completeness scores~\cite{v-measure}. Figure~\ref{fig:v_measure} shows the values for the V measure using the \textsc{KMeans} algorithm on each program representation after ten different runs.

\section{Use Case \#2: Repairing \IPAs}
\label{appendix:repairing}

Table~\ref{tab:num_clusters} presents the number of clusters each clustering method uses for each \IPA. This table was used to generate the cactus plot in Figure~\ref{fig:num_clusters}.

\begin{table}[t!]
\centering
\caption{The number of clusters generated using each clustering approach for each \IPA.}
\label{tab:num_clusters}
\resizebox{\columnwidth}{!}{%
\begin{tabular}{|l|c|c|c|c|c|c|c|}
\hline
\multicolumn{1}{|c|}{\textbf{Exercise}} &
  \textbf{\begin{tabular}[c]{@{}c@{}}\#Correct \\ Submissions\end{tabular}} &
  \textbf{\begin{tabular}[c]{@{}c@{}}Clara \\ Clusters\end{tabular}} &
  \textbf{\begin{tabular}[c]{@{}c@{}}KMeans \\ AAST\end{tabular}} &
  \textbf{\begin{tabular}[c]{@{}c@{}}KMeans \\ AAST+Invs\end{tabular}} &
  \textbf{\begin{tabular}[c]{@{}c@{}}KMeans \\ Invs\end{tabular}} &
  \textbf{\begin{tabular}[c]{@{}c@{}}KMeans \\ Syntax\end{tabular}} &
  \textbf{\begin{tabular}[c]{@{}c@{}}Closest Program \\ (KMeans) \\ AAST+Invs\end{tabular}} \\ \hline
\textbf{lab02/ex01}       & 92 & 18 & 9 & 9 & 9 & 9 & 1 \\ \hline
\textbf{lab02/ex02}       & 84 & 6  & 8 & 8 & 8 & 8 & 1 \\ \hline
\textbf{lab02/ex03}       & 83 & 4  & 8 & 8 & 7 & 8 & 1 \\ \hline
\textbf{lab02/ex04}       & 76 & 20 & 7 & 7 & 7 & 7 & 1 \\ \hline
\textbf{lab02/ex05}       & 80 & 10 & 7 & 7 & 7 & 7 & 1 \\ \hline
\textbf{lab02/ex06}       & 68 & 25 & 6 & 6 & 6 & 6 & 1 \\ \hline
\textbf{lab02/ex07}       & 67 & 17 & 6 & 6 & 6 & 6 & 1 \\ \hline
\textbf{lab02/ex08}       & 49 & 21 & 4 & 4 & 4 & 4 & 1 \\ \hline
\textbf{lab02/ex09}       & 74 & 12 & 7 & 7 & 7 & 7 & 1 \\ \hline
\textbf{lab02/ex10}       & 65 & 17 & 6 & 6 & 6 & 6 & 1 \\ \hline
\textbf{lab03/ex01}       & 70 & 51 & 6 & 6 & 6 & 6 & 1 \\ \hline
\textbf{lab03/ex02}       & 55 & 49 & 5 & 5 & 5 & 5 & 1 \\ \hline
\textbf{lab03/ex03}       & 45 & 27 & 4 & 4 & 4 & 4 & 1 \\ \hline
\textbf{lab03/ex04}       & 28 & 8  & 2 & 2 & 2 & 2 & 1 \\ \hline
\textbf{lab03/ex06}       & 46 & 8  & 4 & 4 & 4 & 4 & 1 \\ \hline
\textbf{lab04/ex01}       & 59 & 32 & 5 & 5 & 5 & 5 & 1 \\ \hline
\textbf{lab04/ex02}       & 47 & 32 & 4 & 4 & 4 & 4 & 1 \\ \hline
\textbf{lab04/ex03}       & 41 & 33 & 4 & 4 & 4 & 4 & 1 \\ \hline
\textbf{lab04/ex08}       & 8  & 6  & 1 & 1 & 1 & 1 & 1 \\ \hline
\textbf{lab05/ex01}       & 4  & 3  & 1 & 1 & 1 & 1 & 1 \\ \hline
\textbf{itsp/lab3/ex2810} & 17 & 9  & 1 & 1 & 1 & 1 & 1 \\ \hline
\textbf{itsp/lab3/ex2811} & 7  & 3  & 1 & 1 & 1 & 1 & 1 \\ \hline
\textbf{itsp/lab3/ex2812} & 17 & 7  & 1 & 1 & 1 & 1 & 1 \\ \hline
\textbf{itsp/lab3/ex2813} & 4  & 4  & 1 & 1 & 1 & 1 & 1 \\ \hline
\textbf{itsp/lab4/ex2824} & 15 & 5  & 1 & 1 & 1 & 1 & 1 \\ \hline
\textbf{itsp/lab4/ex2825} & 10 & 4  & 1 & 1 & 1 & 1 & 1 \\ \hline
\textbf{itsp/lab4/ex2827} & 6  & 6  & 1 & 1 & 1 & 1 & 1 \\ \hline
\textbf{itsp/lab4/ex2831} & 7  & 4  & 1 & 1 & 1 & 1 & 1 \\ \hline
\textbf{itsp/lab4/ex2832} & 17 & 7  & 1 & 1 & 1 & 1 & 1 \\ \hline
\textbf{itsp/lab4/ex2833} & 19 & 9  & 1 & 1 & 1 & 1 & 1 \\ \hline
\textbf{itsp/lab5/ex2865} & 7  & 4  & 1 & 1 & 1 & 1 & 1 \\ \hline
\textbf{itsp/lab5/ex2866} & 11 & 10 & 1 & 1 & 1 & 1 & 1 \\ \hline
\textbf{itsp/lab5/ex2867} & 7  & 4  & 1 & 1 & 1 & 1 & 1 \\ \hline
\textbf{itsp/lab5/ex2868} & 8  & 6  & 1 & 1 & 1 & 1 & 1 \\ \hline
\textbf{itsp/lab5/ex2869} & 7  & 4  & 1 & 1 & 1 & 1 & 1 \\ \hline
\textbf{itsp/lab5/ex2870} & 9  & 8  & 1 & 1 & 1 & 1 & 1 \\ \hline
\textbf{itsp/lab5/ex2871} & 15 & 10 & 1 & 1 & 1 & 1 & 1 \\ \hline
\textbf{itsp/lab6/ex2932} & 3  & 3  & 1 & 1 & 1 & 1 & 1 \\ \hline
\textbf{itsp/lab6/ex2933} & 1  & 1  & 1 & 1 & 1 & 1 & 1 \\ \hline
\textbf{itsp/lab6/ex2936} & 5  & 4  & 1 & 1 & 1 & 1 & 1 \\ \hline
\textbf{itsp/lab6/ex2937} & 2  & 2  & 1 & 1 & 1 & 1 & 1 \\ \hline
\textbf{itsp/lab6/ex2938} & 6  & 4  & 1 & 1 & 1 & 1 & 1 \\ \hline
\textbf{itsp/lab6/ex2939} & 2  & 1  & 1 & 1 & 1 & 1 & 1 \\ \hline
\end{tabular}%
}
\end{table}



\end{document}